\documentclass[12pt]{article}
\usepackage{amsmath}
\begin{document}
\begin{flushright}
KANAZAWA-10-09\\
November, 2010
\end{flushright}
\vspace*{2cm}

\begin{center}
{\Large\bf Dark matter in the supersymmetric radiative seesaw model 
with an anomalous U(1) symmetry}
\vspace*{1cm}

{\Large Daijiro Suematsu}\footnote{e-mail:~suematsu@hep.s.kanazawa-u.ac.jp}
{\Large and Takashi Toma}\footnote{e-mail:~t-toma@hep.s.kanazawa-u.ac.jp}
\vspace*{1cm}\\

{\it Institute for Theoretical Physics, Kanazawa University, 
\\ Kanazawa 920-1192, Japan}
\end{center}
\vspace*{1.5cm} 

\noindent
{\Large\bf Abstract}\\
The existence of an anomalous U(1) symmetry is shown to play a crucial 
role in the supersymmetric radiative seesaw model for neutrino masses.
It explains the smallness of some couplings related to neutrino mass 
generation in a favorable way in addition to cause the hierarchical 
structure of Yukawa couplings of quarks and leptons. 
If it is spontaneously broken to a $Z_2$ subgroup, this $Z_2$ symmetry 
can make a lifetime of the lightest field with its odd parity extremely 
long. Thus, the model has an additional dark matter candidate other 
than the lightest neutralino, which appears in the $R$-parity conserved MSSM.
We discuss the nature of dark matter by taking account of its relation 
to the neutrino mass generation and the lepton flavor violation.

\newpage
\section{Introduction}
The recent astrophysical observations \cite{wmap} and neutrino oscillation 
experiments \cite{oscil} require that the standard model (SM) should be 
extended so as to include dark matter (DM) \cite{susydm} and 
small neutrino masses.
A radiative seesaw model proposed by Ma \cite{Ma:2006km} is a simple and
interesting possibility among such extensions.\footnote{There are a lot of study
for the DM nature in radiative neutrino mass generation models and other 
phenomenological features of such models 
\cite{cdmmeg,lflavor,recon,ext,scdm,fcdm,ncdm}.}  
In this model the existence of DM is intimately related to the 
neutrino mass generation. The combined study of the neutrino oscillation
data,  the lepton flavor violating processes such as $\mu\rightarrow
e\gamma$ and the DM relic abundance can give strong constraints on the
model \cite{cdmmeg,lflavor}. 
On the other hand, the recently reported cosmic ray 
anomaly \cite{pamela,fermi} attracts a lot of attention as a target of
the DM physics. Various work relevant to this has been done on the basis
of the annihilation of DM \cite{recon,mindep,sommerfeld,bwenhance,bw-sty} 
and also the decay of DM \cite{it,decay,gravitino,rparity,hidden}.
If we impose that this neutrino mass model should explain this anomaly, 
it could bring the model additional valuable information 
not only on the mass and the interaction of DM but also
on the lepton flavor structure \cite{bw-sty}. Although the model shows
interesting and promising features on phenomenology, the model cannot 
give any answer to the hierarchy problem unfortunately.
Previous models are usually constructed in the nonsupersymmetric framework.
Thus, the detailed study on the supersymmetrization of the model 
is one of the remaining subjects relevant to the model. 

The supersymmetric extension of the model has been considered in
\cite{sma,fks}.\footnote{A relevant supersymmetric model is considered
in a different context in \cite{e6}.}
Two DM candidates appear in these extensions as long as $R$-parity 
conservation is assumed. The existence of two kinds of $Z_2$ symmetry,
that is, $R$-parity and $Z_2$ which forbids tree-level neutrino masses,
guarantees the stability of these two DM candidates.
In that case both of them contribute to the expected DM relic 
abundance.\footnote{Multi-component DM models have been considered in several
contexts \cite{multidm}.}
However, if one of them is unstable but has an longer life time than the
age of universe, this decaying DM may cause the additional contribution 
cosmic ray, which brings anomalous excess of the charged particle flux 
over the expected background. Such a possibility has been studied 
in a supersymmetrized model in \cite{fks}, where $Z_2$ is assumed to 
be weakly broken by the anomaly effect. 
Since the anomaly induced interaction is strongly suppressed,
it causes the DM decay but its life time can be long enough.

In this paper, we consider a modification of the original one given in 
\cite{fks} by introducing the an anomalous U(1) gauge symmetry.
Spontaneous breaking of this symmetry causes the weakly broken $Z_2$ symmetry 
at low energy regions naturally in addition to the conserved $R$-parity. 
It makes us possible to understand the smallness of
some couplings through the Froggatt-Nielsen mechanism \cite{fn,yh},
which is required to generate the small neutrino masses.\footnote{
An anomalous U(1) symmetry has been considered in various
phenomenological contexts \cite{chris}.
In a radiative seesaw mechanism, it is also considered 
in \cite{srz}. However, the present model is different from it in 
the nature of dark matter and also the $R$-parity conservation.} 
Simultaneously, hierarchical masses of quarks and charged leptons
are also explained by the same origin.
The model also shows another interesting features relevant to the DM
phenomenology. 
We give its analysis which includes the cosmic ray anomaly
expected in this model, the characteristic gamma predicted
in the DM radiative decay and the direct search of the stable 
DM through the elastic scattering with nuclei.

The paper is organized as follows. In section 2 we give the low energy
effective model with an anomalous U(1) gauge symmetry. 
Spontaneous breaking of this symmetry gives the $Z_2$ symmetry which is
relevant to the neutrino mass generation. 
The neutrino mass and mixing, the lepton flavor
violating processes and the DM relic abundance are studied in this framework.
In section 3 the nature of DM is discussed from a view point of both
direct and indirect searches.
Section 4 is devoted to the summary.
  
\section{A supersymmetric extension}
The radiative seesaw model proposed in \cite{Ma:2006km} is an extension of the
SM with three right-handed neutrinos and an inert doublet 
scalar.\footnote{The origin of the second doublet is discussed in
\cite{second}.}
 It has no coupling with quarks and no vacuum expectation value (VEV).
Although the model has interesting phenomenological features 
as discussed in \cite{cdmmeg,lflavor,ext},
it can not give an answer to the hierarchy problem. 
Thus, it is worthy to consider the supersymmetric extension 
of the model and to study phenomenology associated to it. 
The supersymmetrization of the model requires to 
introduce inert doublet chiral superfields $\eta_u$ and 
$\eta_d$ and also a singlet chiral superfield $\phi$
to the minimal supersymmetric SM (MSSM) \cite{fks}.
They play the similar role as the inert doublet scalar in the original
nonsupersymmetric model and bring required terms in the Lagrangian for
the neutrino mass generation. As the origin of the $Z_2$ symmetry which
forbids tree level neutrino masses,
we suppose the existence of an anomalous U(1)$_X$ 
gauge symmetry. We assume that the $R$-parity is conserved. 
Matter contents of the model and their relevant quantum numbers are 
summarized in Table 1. 

\begin{table}[t]
\footnotesize
\hspace{1.7cm}\begin{tabular}{|c|ccccccc|} \hline
$\Psi$ &$Q_i$ & $\bar U_i$ & $\bar D_i$ & $L_i$ & $\bar E_i$
	       &$H_u$&$H_d$ \\ \hline
SU(2)$_L\times$ U(1)$_Y$ & $(\underline{2},\frac{1}{6})$ &
$(\underline{1},\frac{-2}{3})$ & $(\underline{1},\frac{1}{3})$ &
$(\underline{2},\frac{-1}{2})$ & $(\underline{1}, 1)$ &
$(\underline{2},\frac{1}{2})$ & $(\underline{2},\frac{-1}{2})$ \\ \hline
$R$ & $-$ & $-$ &$-$& $-$ & $-$ & $+$&  $+$  \\ \hline
$X$ 
 &~ $2n_{Q_i}$~ &~ $2n_{U_i}$~ &~ $2n_{D_i}$~ &~ $2n_{L_i}$~ &~ $2n_{E_i}$~  
&~ $2n_{H_u}$~ &~ $2n_{H_d}$  
\\ \hline
$Z_2$ & $+$ & $+$ &$+$& $+$ & $+$ &  $+$&  $+$  \\ \hline
\end{tabular}

\hspace{1.7cm}\begin{tabular}{|c|cccccc|} \hline
$\Psi$ & $\bar N_i $ & $\eta_u $ & $\eta_d$ & $\phi $ & $\Sigma_+$
	       &$\Sigma_-$ 
\\ \hline
SU(2)$_L\times$ U(1)$_Y$ & $(\underline{1}, 0)$ &
$(\underline{2},\frac{1}{2})$ & $(\underline{1},\frac{-1}{2})$ &
$(\underline{1}, 0)$ & $(\underline{1}, 0)$ &
$(\underline{1}, 0)$ \\ \hline
$R$  &  $+$ &  $-$ & $-$& $-$ &$+$ & $+$  \\ \hline
$X$  &~ $2n_{N_i}+1$~  & $2n_{\eta_u}+1$~ & $2n_{\eta_d}+1$~ & $2n_\phi +1$~ 
&~~ $2n_+$~~ &~ $-2$~~  
\\ \hline
$Z_2$ & $-$ &  $-$ & $-$& $-$ &$+$ & $+$  \\ \hline
\end{tabular}
\begin{center}
\caption{\footnotesize Matter contents and their quantum number. 
$X$ represents the charge of the anomalous U(1)$_X$ and 
each $n_\Psi$ is an integer.
$Z_2$ is a remnant symmetry of U(1)$_X$ caused by the
symmetry breaking due to $\langle\Sigma_\pm\rangle\not=0$.}
\end{center}
\end{table}
\normalsize

\subsection{A low energy effective model}
First, we discuss spontaneous breaking of the anomalous U(1)$_X$ symmetry 
at a high energy scale, which brings the low energy effective theory.
Hierarchical couplings and masses are found to be generated 
in the low energy theory through this breaking.
The vacuum is expected to be determined as the flat direction of the 
$D$-term of U(1)$_X$. 
The $D$-term for relevant hidden fields is given by 
\begin{equation}
V_D=\frac{g_X^2}{2}\Big[(2n_\phi+1)|\phi|^2 + 2n_+|\Sigma_+|^2 -2|\Sigma_-|^2 
+\xi_X\Big]^2.
\end{equation} 
$\xi_X$ is the U(1)$_X$ Fayet-Iliopoulos $D$-term.
It is expressed in string models as \cite{fi}
\begin{equation}
\xi_X=\frac{{\rm Tr}X}{192\pi^2}g_X^2M_{\rm pl}^2
\equiv \delta_{\rm GS}~g_X^2M_{\rm pl}^2,
\end{equation}
where $M_{\rm pl}$ is the reduced Planck mass and 
$X$ stands for the anomalous U(1)$_X$ charge of the fields.
On the other hand, the lowest order superpotential for these 
fields can be written as
\begin{equation}
W_h=\frac{c_+}{M_{\rm pl}^{n_+-2}}\Sigma_+\Sigma_-^{n_+}
+\frac{c_\phi}{M_{\rm pl}^{2n_\phi}}\phi^2\Sigma_-^{2n_\phi+1},
\end{equation}
where $c_+$ and $c_\phi$ are considered as real constants.
If we suppose supersymmetry breaking in the hidden sector,
supersymmetry breaking terms appear in the scalar potential
of the hidden sector. 
The scalar potential derived from $W_h$ may be written as
\begin{eqnarray}
V_F&=&\frac{c_+^2}{M_{\rm pl}^{2(n_+-2)}}|\Sigma_-|^{2n_+}
+\frac{4c_\phi^2}{M_{\rm pl}^{4n_\phi}}|\phi\Sigma_-^{2n_\phi+1}|^2
+\frac{c_+^2n_+^2}{M_{\rm pl}^{2(n_+-2)}}|\Sigma_+\Sigma_-^{n_+-1}|^2
\nonumber \\
&&+\frac{c_\phi^2(2n_\phi+1)^2}{M_{\rm pl}^{4n_\phi}}
|\phi^2\Sigma_-^{2n_\phi}|^2 
+\left(\frac{c_+n_+c_\phi(2n_\phi+1)}{M_{\rm pl}^{n_++2n_\phi-2}}
(\Sigma_+\Sigma_-^{n_+-1})^\ast\phi^2\Sigma_-^{2n_\phi}+{\rm
h.c.}\right), \nonumber \\
&&-F_1\Sigma_-^2 - F_2\Sigma_+^2,
\end{eqnarray}
where $F_1$ and $F_2$ represent the VEVs of $F$-components of some chiral
superfields in the hidden sector. They bring the supersymmetry breaking of
$O(10^{11})$~GeV in the hidden sector and also induce the soft terms
of $O(1)$~TeV through the gravity mediation in the observable sector.

Here we assume that the VEVs of $\Sigma_\pm$ and $\phi$ are real, for
simplicity.
Then, the minimization of the potential $V=V_D+V_F$ along the $D$-flat
direction suggests that there exists a vacuum defined by
\begin{equation}
\langle\Sigma_-\rangle\simeq \sqrt{\frac{\xi_X}{2}}\gg 
\langle\Sigma_+\rangle, \qquad
\langle\phi\rangle=0,
\label{vev}
\end{equation} 
as long as $c_+$ is sufficiently suppressed.
For example, if ${\rm Tr}X\sim 150/g_X^2$ is satisfied, 
$\langle\Sigma_-\rangle\sim 0.2M_{\rm pl}$ is expected.
Moreover, $\langle\Sigma_+\rangle\simeq 10^{-4}M_{\rm pl}$ is 
also expected for sufficiently suppressed values of $c_+$
such as $O(10^{-7})$.
Although this kind of vacuum can be realized only for the 
finely tuned parameters, we assume it in the following study. 

If we adopt this vacuum to fix the low energy effective theory, 
the superpotential invariant under the imposed symmetry is obtained 
with the effectively induced parameters such as 
\begin{eqnarray}
W&=&h_{ij}^U Q_{i} \bar U_{j}  H_u
+ h_{ij}^DQ_{i} \bar D_{j} H_d
+ h_{i}^EL_{i} \bar E_{i} H_d
+\mu_H H_u H_d, \nonumber \\
&+&h_{ij}^NL_{i}\bar N_{j} \eta_u
+\lambda_u\eta_u H_d \phi
+\lambda_d\eta_d H_u \phi
+\mu_\eta\eta_u\eta_d
+ \frac{1}{2}M_i\bar N_i\bar N_i 
+\frac{1}{2}\mu_\phi\phi^2.
\label{superpot}
\end{eqnarray}
The invariance of each term under $R\times$U(1)$_X$ is guaranteed since
the effective couplings and masses are generated through the VEVs
$\langle\Sigma_\pm\rangle$ as shown below (see also Appendix A).
The MSSM superpotential is contained in the first line, 
while the second line includes additional terms to the MSSM.
Yukawa couplings for the charged leptons and the mass matrix 
for the right-handed neutrinos are supposed to be flavor 
diagonal.\footnote{This can be justified as long as the relevant
original couplings $y_{ijk}$ and $y_{ij}$ in eq.~(\ref{effp}) 
are flavor diagonal. }

The effective parameters in this superpotential are induced from higher
order invariant interaction terms which contain a suitable number of 
$\Sigma_\pm$. 
The dominant contribution are determined by the lowest order
term of the following form:
\begin{eqnarray}
&&h_{ijk}=y_{ijk}\left(\frac{\langle\Sigma_\pm\rangle}{M_{\rm
		  pl}}\right)^{n_{ijk}}, \quad
n_{ijk}=-\frac{X_i+X_j+X_k}{X_{\Sigma_\pm}}\quad 
{\rm for}~h_{ijk}\Psi_i\Psi_j\Psi_k, \nonumber \\
&&\mu_{ij}=y_{ij}M_{\rm pl}
\left(\frac{\langle\Sigma_\pm\rangle}{M_{\rm pl}}\right)^{n_{ij}}, \quad
n_{ij}=-\frac{X_i+X_j}{X_{\Sigma_\pm}} \quad 
{\rm for}~\mu_{ij}\Psi_i\Psi_j.
\label{effp} 
\end{eqnarray}
The original coupling constants $y_{ijk}$ and $y_{ij}$ in the 
nonrenormalizable interaction terms are considered to be values of $O(1)$.
If the singlet scalars $\Sigma_\pm$ obtain the VEVs as discussed above, 
these VEVs cause hierarchical structure in the Yukawa couplings
of quarks and leptons, and also suppress several parameters in the
superpotential as found from eqs.~(\ref{vev}) and (\ref{effp}).
In fact, as long as suitable U(1)$_X$ charges are assigned to 
quarks and leptons, we find that the hierarchical mass eigenvalues and 
mixing are generated via Froggatt-Nielsen mechanism \cite{fn}.
In the similar way, the Yukawa couplings $\lambda_{u,d}$ are 
largely suppressed, and also 
$\mu_\eta$ and $M_i$ can take values of $O(1)$~TeV.
These parameters are relevant to the neutrino mass generation 
and their values can be favorable for it as seen in the next part.
We give more detailed discussion by giving such concrete examples 
for the charge assignment in the Appendix A.
We also find that $\langle\Sigma_\pm\rangle$ breaks the anomalous 
U(1)$_X$ symmetry spontaneously to its $Z_2$ subgroup.  
This is obvious from the charge assignment shown in Table 1. 
This $Z_2$ symmetry plays a crucial role in the DM phenomenology.

Here we also note an interesting point related to the anomaly 
induced interaction. 
We have $L_i\eta_u$ as an only renormalizable 
operator which breaks the U(1)$_X$ invariance but is invariant under 
the SM gauge symmetry and the $R$-parity.
It is not included in the low energy superpotential (\ref{superpot}) 
since we can not make it U(1)$_X$ invariant by multiplying 
any number of $\Sigma_\pm$. 
However, it can be U(1)$_X$ invariant if it is accompanied by a 
dilaton chiral superfield $S$.
As is well known, if the anomalous U(1)$_X$ gauge transformation 
$V_X\rightarrow V_X +i(\Lambda-\Lambda^\dagger)/2$ is associated with the 
shift of the dilaton field such as 
\begin{equation}
S\rightarrow S+i\delta_{\rm GS}\Lambda,
\end{equation} 
the anomaly cancellation for U(1)$_X$ is completed \cite{gs,adm}.
Noting this fact, we observe that the following non-perturbative
superpotential is also invariant under the imposed 
symmetry \cite{adm}:\footnote{The effect of the appearance of this term has 
been studied in the case of discrete symmetry in \cite{fks}.}
\begin{eqnarray}
&&W_{\rm np}=c_iM_{\rm pl}e^{-b_i}L_i\eta_u, \nonumber \\
&&b_i=\frac{(2n_i^L+2n_{\eta_u}+1)}{\delta_{\rm GS}}S
\sim \frac{192\pi^2(2n_i^L+2n_{\eta_u}+1)}{\rm Tr X}
\label{anom}
\end{eqnarray}
where we use $\langle S\rangle$ and $c_i$ are constants
of $O(1)$. 
The expression for $b_i$ shows that it is determined by the U(1)$_X$ 
charge of the field contents of the model including the hidden sector.
We also note that this term breaks the remnant $Z_2$ symmetry 
very weakly. The scale of its violation is determined by both the value 
of ${\rm Tr}X$ and the charges of $L_i$ and $\eta_u$. 

Soft supersymmetry breaking terms associated with 
the superpotential $W$ and $W_{\rm np}$ are introduced as follows,
\begin{eqnarray}
{\cal L}_{SB}
&=&-\tilde m_{\eta_u}^2\tilde{\eta}_u^\dag\tilde{\eta}_u-
\tilde m_{\eta_d}^2\tilde{\eta}_d^\dag\tilde{\eta}_d
-\tilde m_{\bar N}^2 \tilde{\bar N}_i^\dag\tilde{\bar N}_i 
-\tilde m_{\phi}^2 \tilde{\phi}^\dag\tilde\phi \nonumber \\
&&+A(h_{ij}^N\tilde L_i\tilde{\bar N}_j\tilde\eta_u 
+\lambda_u \tilde{\eta}_u {H}_d\tilde\phi+
 \lambda_d \tilde{\eta}_d{H}_u\tilde\phi + {\rm h.c.}) \nonumber \\
&&-B\left(\mu_HH_uH_d+\mu_\eta\tilde{\eta}_u \tilde{\eta}_d+ 
\frac{1}{2}\mu_\phi\tilde{\phi}^2+\frac{1}{2}M_i \tilde{\bar N}_i^{2}
+c_iM_{\rm pl}e^{-b_i}\tilde{L}_i\tilde{\eta}_u 
+{\rm h.c.}\right),
\label{softsb}
\end{eqnarray}
where the additional part to the MSSM is listed alone.
The scalar component is represented by putting a tilde on the 
character of the corresponding chiral superfield except
for the ordinary Higgs chiral superfields $H_u$ and $H_d$. 
Universality of the soft supersymmetry breaking $A$- and $B$-parameters 
is assumed to be satisfied here. 
Moreover, we confine our consideration to the case in which 
the soft scalar masses for all scalar partners of quarks and leptons
are flavor diagonal and universal.
They are denoted by $m_0$ in the following analysis and we
assume $A=B=m_0$, for simplicity.

We should remind the reader that the soft scalar masses have the
contributions from the anomalous U(1)$_X$ $D$-term such as
\begin{equation}
\Delta \tilde m_i^2=g_X^2X_i\left(C_h+2n_{H_u}\langle H_u^0\rangle^2
+2n_{H_d}\langle H_d^0\rangle^2\right),
\end{equation}
where $C_h$ stands for the remnant contribution from the hidden sector.
Although these are flavor dependent, we confine our study to the case
$m_0^2 \gg \Delta \tilde m_i^2$, which may be realized on the
vacuum with a finely tuned $C_h$.
In such a case there is no new dangerous origin for the flavor changing neutral 
current (FCNC) caused by the soft supersymmetry breaking terms.
In the following study the effective parameters in eq.~(\ref{superpot})
 and also the soft supersymmetry breaking parameters in
eq.~(\ref{softsb}) are treated as real except for $h^N_{ij}$, for simplicity.

\subsection{Neutrino mass and mixing}
Neutrino masses are generated through the one-loop diagram shown in
Fig.~1 when the Higgs doublet scalars $H_u^0$ and $H_d^0$ obtain the VEVs.
As found from eq.~(\ref{fnco}), 
since $\lambda_u$ and $\lambda_d$ are very small, the mixing between
$(\tilde\eta_u^{0\ast}, \tilde\eta_d^0)$ and
$(\tilde\phi^\ast,\tilde\phi)$ can be treated as an insertion in the
calculation of these diagrams with good accuracy.
The mass matrices for $(\tilde\eta_u^{0\ast}, \tilde\eta_d^0)$ and 
$(\tilde\phi^\ast,\tilde\phi)$ are written as
\begin{equation}
{\cal M}_{\eta^0}=\left(\begin{array}{cc}
\bar m_{\eta_u}^2 & B\mu_\eta \\
 B\mu_\eta & \bar m_{\eta_d}^2 \\
\end{array} \right), \qquad
{\cal M}_\phi=\frac{1}{2}\left(\begin{array}{cc}
\bar m_{\phi}^2 & B\mu_\phi \\
 B\mu_\phi & \bar m_\phi^2 \\
\end{array} \right)
\end{equation}
where $\bar{m}_{\eta_{u,d}}^2\simeq m^2_0
+\mu_\eta^2+\lambda_{u,d}^2v_{d,u}^2$ and 
$\bar{m}_\phi^2\simeq m^2_0+\mu_\phi^2+
\lambda_u^2v_d^2+\lambda_d^2v_u^2$.
If we define the mass eigenstates of these mass matrices by
\begin{equation}
\left(\begin{array}{c}\tilde\eta_+ \\ \tilde\eta_- \end{array}\right)=
\left(\begin{array}{cc}
\cos\theta_\eta & \sin\theta_\eta \\ -\sin\theta_\eta & \cos\theta_\eta \\
\end{array}\right)
\left(\begin{array}{c}\tilde\eta_u^{0\ast} \\ \tilde\eta_d^0 \end{array}\right),
\quad
\left(\begin{array}{c}\tilde\phi_+ \\ \tilde\phi_- \end{array}\right)=
\left(\begin{array}{cc}
\cos\theta_\phi & \sin\theta_\phi \\ -\sin\theta_\phi & \cos\theta_\phi \\
\end{array}\right)
\left(\begin{array}{c}\tilde\phi^\ast \\ \tilde\phi \end{array}\right),
\end{equation}
the mass eigenvalues and the mixing angles can be written as 
\begin{eqnarray}
&&m_{\eta\pm}^2=\frac{1}{2}\left(\bar m_{\eta_u}^2+\bar m_{\eta_d}^2
\pm\sqrt{(\bar m_{\eta_u}^2-\bar m_{\eta_d}^2)^2+4B^2\mu_\eta^2}~ 
\right), 
\quad \tan 2\theta_\eta=\frac{2B\mu_\eta}{\bar m_{\eta_d}^2-\bar
 m_{\eta_u}^2}, \nonumber \\
&&m_{\phi\pm}^2=\bar m_\phi^2 \pm B\mu_\phi, 
\qquad \theta_\phi =\frac{\pi}{4}.
\label{etamass}
\end{eqnarray}
We find that the scalar superpartners $(\tilde{\bar
N}_i^\ast,\tilde{\bar N}_i)$ of
$\bar N_i$ satisfy the same relations as the ones of 
$(\tilde\phi^\ast, \tilde\phi)$ as shown above.
Their mass eigenvalues $M_{i\pm}^2$ can be read off from the expression for
$m_{\phi\pm}^2$ by replacing $\mu_\phi$ and
$\bar m_\phi^2$ with $M_i$ and $m_0^2+M_i^2$, respectively.

We can calculate the neutrino masses generated through the one-loop
diagrams by using these. Since $\mu_\phi^2$ is expected to be larger than
$\mu_\eta^2$ and $m_0^2$, we find that the dominant contribution is
caused by the diagram (a) in Fig.~1 and it is estimated as
\begin{eqnarray}
({\cal M}_\nu)_{\alpha\beta}&=&
\frac{\lambda_u\lambda_dv_uv_d\sin 2\theta_\eta}{16\pi^2 }
\sum_{i=1}^3h_{\alpha i}h_{\beta i}M_i
\Big(g(M_i,m_{\eta +})\cos^2\theta_\eta
-g(M_i,m_{\eta -})\sin^2\theta_\eta \nonumber \\
&&-f(M_i,m_{\eta +},m_{\eta -})\cos 2\theta_\eta\Big),
\label{nmtr}
\end{eqnarray}
where $f$ and $g$ are defined as
\begin{eqnarray}
&&f(m_a,m_b,m_c)=\frac{m_a^2m_b^2\ln(m_b^2/m_a^2)
+m_b^2m_c^2\ln(m_c^2/m_b^2)+
m_c^2m_a^2\ln(m_a^2/m_c^2)}{(m_c^2-m_a^2)(m_a^2-m_b^2)(m_b^2-m_c^2)},
\nonumber \\
&&g(m_a,m_b)=\frac{m_b^2 -m_a^2+m_a^2\ln(m_a^2/m_b^2)}{(m_b^2-m_a^2)^2}.
 \label{mnu}
\end{eqnarray}
Other two diagrams (b) and (c) with the component of $\phi$ as 
an internal line are expected to bring the 
subdominant contributions. We give the explicit
expressions of their contributions in the Appendix B.
The universal soft supersymmetry breakings make the situation simple since
$\theta_\eta=\pi/4$ is satisfied.
If $M_i$ and $\bar m_{\eta_{u,d}}$ have the values of $O(1)$~TeV,
the mass eigenvalues of neutrinos must be controlled by a small
parameter $\lambda_u\lambda_d$ of $O(10^{-8})$. 
As long as the anomalous U(1)$_X$ charge is assigned suitably, these can
be naturally realized as found from the examples given in the Appendix A.

\input epsf
\begin{figure}[t]
\begin{center}
\epsfxsize=12cm
\leavevmode
\epsfbox{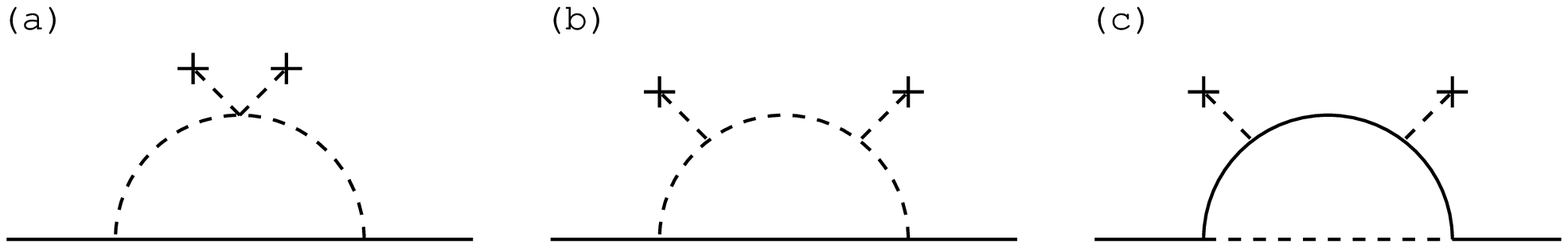}
\end{center}

\vspace*{-3mm}
{\footnotesize {\bf Fig.~1}~~One-loop diagrams contributing to the
 neutrino masses. Two diagrams (b) and (c) have an internal line of
 the scalar $\tilde\phi$ or its fermionic partner, respectively.}
\end{figure}

Here we focus our attention to the flavor structure of the neutrino Yukawa 
couplings\footnote{This structure is chosen adhoc here since
it has several interesting features as shown below.} \cite{lflavor,recon}
\begin{equation}
h_{ei}^N=0, \quad h_{\mu i}^N=h_{\tau i}^N\equiv h_i
 ~(i=1,2), \qquad
h_{e3}^N=h_{\mu 3}^N=-h_{\tau 3}^N\equiv h_3.
\label{yukawa}
\end{equation}
This flavor structure is consistent with the U(1)$_X$ invariance 
as long as all of $y^N_{ij}$ take the same values except for 
$y^N_{ei}$, which should be assumed to be $y^N_{ei}=0$ for $i=1,2$ (see
Appendix A). 
An interesting point of this flavor structure is that the neutrino mass 
matrix in eq.(\ref{nmtr}) takes the following simple form:
\begin{equation}
{\cal M}_\nu=\left(
\begin{array}{ccc}
0 & 0 & 0\\ 0 & 1 & 1 \\ 0 & 1 & 1 \\ \end{array}\right)
(h_{1}^2\Lambda_1+h_{2}^2\Lambda_2)
+\left(
\begin{array}{ccc}
1 & 1 & -1\\ 1 & 1 & -1 \\ -1 & -1 & 1 \\ \end{array}\right)
h_{3}^2\Lambda_3,
\label{nmass}
\end{equation}
where $\Lambda_i$ fixes the mass scale for the neutrino masses 
as follows,\footnote{If the mass matrix is modified from eq.~(\ref{nmtr})
due to other contributions neglected as the subdominant ones here, 
the ambiguity is confined into the $\Lambda_i$. The MNS matrix is not
affected as long as the condition (\ref{yukawa}) is satisfied.}
\begin{equation}
\Lambda_i=\frac{\bar\lambda v^2M_i}
{32\pi^2}\Big(g(M_i,m_{\eta +})-g(M_i,m_{\eta -})\Big), \qquad
\bar\lambda\equiv\frac{\lambda_u\lambda_d\tan\beta}{1+\tan^2\beta}, 
\label{mscale}
\end{equation}
where $\sin\beta=v_u/v$ and $\cos\beta=v_d/v$.
This mass matrix automatically derives the tri-bimaximal mixing, which is 
favored by the neutrino oscillation data.  
In fact, it is easily checked that the MNS matrix for this neutrino mass
model is given by
\begin{equation}
U_{MNS}=\left(\begin{array}{ccc}
\frac{2}{\sqrt 6} & \frac{1}{\sqrt 3} & 0\\
 \frac{-1}{\sqrt 6} & \frac{1}{\sqrt 3} & \frac{1}{\sqrt 2}\\
\frac{1}{\sqrt 6} & \frac{-1}{\sqrt 3} & \frac{1}{\sqrt 2}\\
\end{array}\right)
\left(\begin{array}{ccc}
1 &0 & 0\\
0 & e^{i\alpha_1} & 0 \\
0 & 0 & e^{i\alpha_2} \\
\end{array}\right), 
\end{equation}
where Majorana phases $\alpha_{1,2}$ are expressed as
\begin{equation}
\alpha_1=\varphi_3, \qquad
\alpha_2=\frac{1}{2}\tan^{-1}
\left(\frac{|h_1|^2\Lambda_1\sin 2\varphi_1+
|h_2|^2\Lambda_2\sin 2\varphi_2}
{|h_1|^2\Lambda_1\cos 2\varphi_1+|h_2|^2\Lambda_2\cos 2\varphi_2}\right)
\end{equation}
by using $\varphi_i={\rm arg}(h_i)$.
Here it should be reminded that Majorana phases $\alpha_{1,2}$ do not affect 
the neutrino oscillations.

Taking account of the nature discussed above and also the fact that one of 
the eigenvalues of the mass matrix (\ref{nmass}) is zero,
the remaining mass eigenvalues are found to be equal to 
$\sqrt{\Delta m^2_{\rm atm}}$ and $\sqrt{\Delta m_{\rm sol}^2}$. 
This is required for the explanation of neutrino oscillation data. 
Thus, we find that the model parameters should satisfy the relations 
\begin{equation}
|h_{1}^2\Lambda_1+h_{2}^2\Lambda_2|\simeq 
\frac{\sqrt{\Delta m_{\rm atm}^2}}{2}, \qquad
|h_{3}^2\Lambda_3|\simeq \frac{\sqrt{\Delta m_{\rm sol}^2}}{3}.
\label{c-oscil}
\end{equation}
The neutrino Yukawa couplings $h_i$ and the right-handed neutrino 
masses $M_i$ should correlate each other so as to satisfy these relations.
Phenomenological study of the model should be proceeded under these
constraints. 
In the following discussion, we restrict our study 
to the case with $M_1~{^<_\sim}~M_2<M_3$,
 which allows us to take $\Lambda_1\simeq\Lambda_2$.\footnote{
It may be useful to note that 
this case has some advantages as discussed in \cite{recon,bw-sty},
but not only for simplicity.} 
Thus, free parameters relevant to the
analysis of DM phenomenology are summarized as
\begin{equation}
M_1, \quad M_3, \quad \bar\lambda, \quad \mu_\eta, \quad  m_0.
\end{equation}
If we suppose fixed the values for $M_i$, $\mu_\eta$ and $\tilde m_0$, 
the conditions (\ref{c-oscil})
determine a value of $\bar\lambda\sqrt{|h_1^2+h_2^2|}$ and 
$\bar\lambda|h_3|$. A larger $\bar\lambda$ gives a smaller 
value of neutrino Yukawa couplings $\bar\lambda\sqrt{|h_1^2+h_2^2|}$ and 
$\bar\lambda|h_3|$. This feature becomes crucial when the constraints
from lepton flavor violating processes and DM relic abundance
are taken into account.

\subsection{Constraints from lepton flavor violating processes}
Since we suppose the slepton mass matrix is flavor
diagonal and universal, there are no new FCNC source in the
slepton sector. The FCNC is induced only through the MNS matrix 
elements which appear in the Higgsino exchange diagrams.
Thus, if Higgsinos are much heavier than gauginos, the ordinary 
contributions to the FCNC caused by the supersymmetric partners 
are sufficiently suppressed in this model.
On the other hand, the extension for the neutrino mass generation induces
the one-loop contribution to the lepton flavor 
violating processes such as $\mu\rightarrow e\gamma$ as in the
nonsupersymmetric case. Their diagrams are shown in Fig.~2.
Since these diagrams do not need the mixing between $\eta_u$ and
$\eta_d$ unlike the case of neutrino masses (see Fig.~1), it causes 
the large contributions to these processes. 
In that case the dominant contributions to the lepton flavor violating
processes are given by these diagrams.  
They give the constraints on the model, which is the different 
from the ones in the MSSM.
These processes may be used as the probe of the model in the future 
experiments. 

\begin{figure}[t]
\begin{center}
\epsfxsize=10cm
\leavevmode
\epsfbox{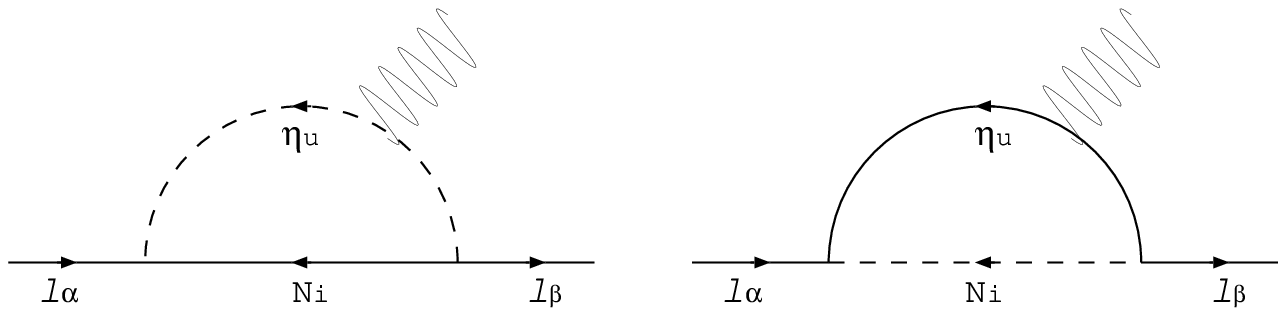}
\vspace*{-3mm}

{\footnotesize {\bf Fig.~2}~~One-loop diagrams causing the lepton flavor
 violating processes $\ell_\alpha\rightarrow\ell_\beta\gamma$.}
\end{center}
\end{figure}

Branching ratio of the lepton flavor violating process 
$\ell_\alpha\rightarrow\ell_\beta\gamma$ is given as
\begin{eqnarray}
&&Br(\ell^-_\alpha\rightarrow\ell^-_\beta\gamma)
=\frac{3\alpha}{64\pi}
\left|\sum_{i=1}^3h_{\alpha i}h_{\beta i}^\ast
\sum_{a=\pm}\left[\frac{1}{2G_Fm_{\eta a}^2}
F_2\left(\frac{M_{i}^2}{m_{\eta a}^2}\right)
+\frac{a}{2G_F\mu_\eta^2}F_2\left(\frac{M_{i a}^2}{\mu_\eta^2}\right)
\right]\right|^2 \nonumber \\
&&\hspace*{3cm} 
\times Br(\ell^-_\alpha\rightarrow\ell^-_\beta\bar\nu_\beta\nu_\alpha),
\end{eqnarray}
where $m_{\eta a}$ and $M_{i a}$ are the mass eigenvalues defined 
in eq.~(\ref{etamass}) and the statements below it respectively. 
The function $F_2(x)$ is defined as
\begin{equation}
F_2(x)=\frac{1-6x+3x^2+2x^3-6x^2\ln x}{6(1-x)^4}.
\end{equation}
If we use the assumed flavor structure for the neutrino Yukawa 
couplings (\ref{yukawa}), we find that
\begin{eqnarray}
Br(\mu\rightarrow e\gamma)&\simeq&
\frac{3\alpha |h_{3}|^4}{64\pi}
\left[\sum_{a=\pm}\left\{\frac{1}{2G_Fm_{\eta a}^2}
F_2\left(\frac{M_3^2}{m_{\eta a}^2}\right)
+\frac{a}{2G_F\mu_\eta^2}
F_2\left(\frac{M_{3a}^2}{\mu_\eta^2}\right)\right\}\right]^2, \nonumber \\
Br(\tau\rightarrow \mu\gamma)&\simeq&
\frac{0.51\alpha}{64\pi}\left[
\sum_{a=\pm}\left\{\frac{1}{2G_Fm_{\eta a}^2}
\left((|h_{1}|^2+|h_{2}|^2)F_2
\left(\frac{M_1^2}{m_{\eta a}^2}\right)
-|h_{3}|^2F_2\left(\frac{M_3^2}{m_{\eta a}^2}\right)\right)\right.\right.
\nonumber \\
&&\left.\left.+\frac{a}{2G_F\mu_\eta^2}\left(
(|h_{1}|^2+|h_{2}|^2)F_2\left(\frac{M_{1a}^2}{\mu_\eta^2}\right)
-|h_{3}|^2F_2\left(\frac{M_{3a}^2}{\mu_\eta^2}\right)\right)\right\}\right]^2
\label{eqlfv}
\end{eqnarray}

\begin{figure}[t]
\begin{center}
\epsfxsize=7cm
\leavevmode
\epsfbox{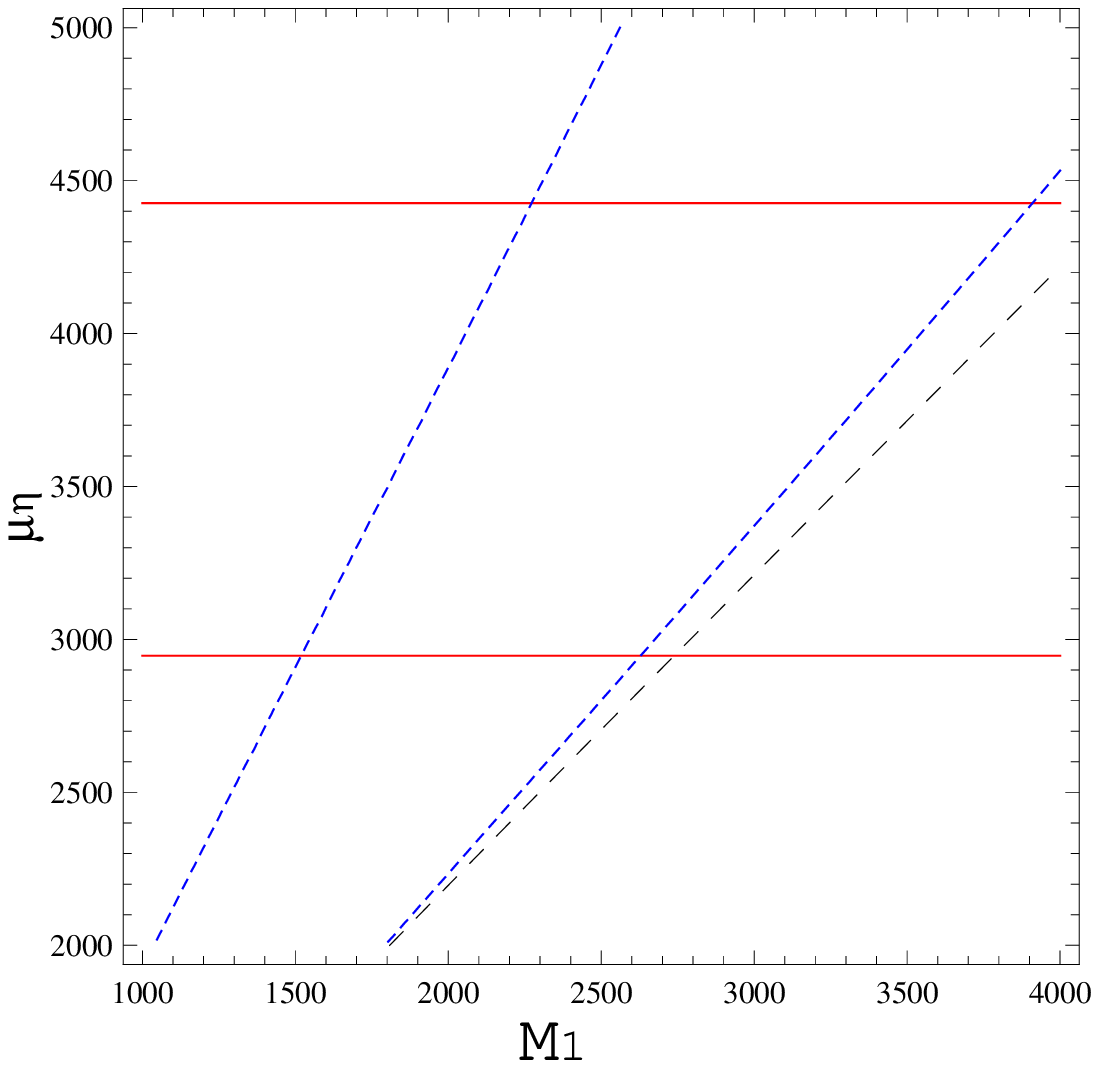}
\hspace*{7mm}
\epsfxsize=8cm
\leavevmode
\epsfbox{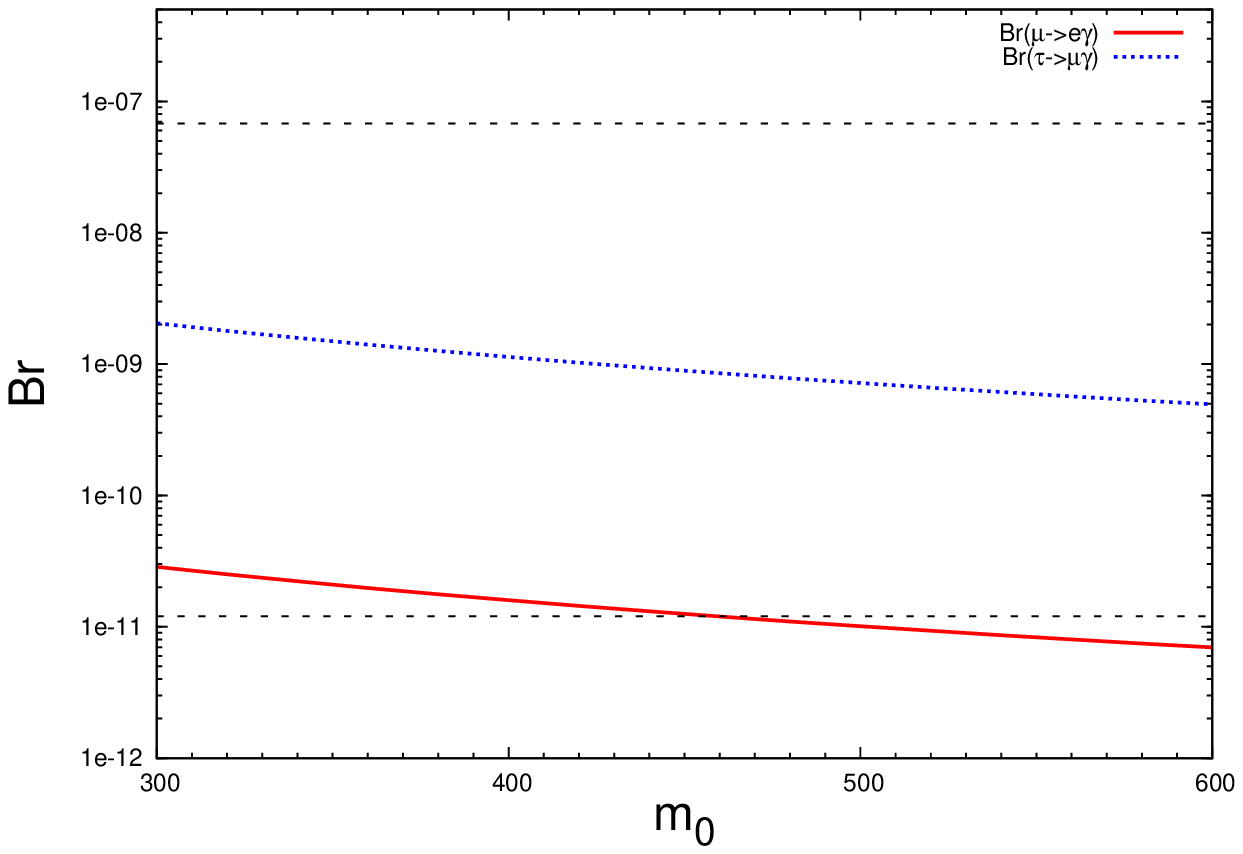}
\end{center}
\vspace*{-3mm}
{\footnotesize {\bf Fig.~3}~~The left frame shows the contours for 
$Br(\mu\rightarrow e\gamma)$ (red solid lines) and 
$Br(\tau\rightarrow\mu\gamma)$ (blue dashed lines)
under the conditions (\ref{c-oscil}) imposed by the neutrino oscillation 
data. A thin black dashed line represents a line for $m_{\eta-}=M_1$.
The right frame shows the $m_0$ dependence of each branching
 ratio. Thin black dotted lines represent the present
 experimental bounds for $\mu\rightarrow e\gamma$ and 
$\tau\rightarrow\mu\gamma$, respectively. We use a GeV unit for $M_1$,
 $m_0$ and $\mu_\eta$.}
\end{figure}

Now we examine the parameter regions consistent with both the neutrino
oscillation data and the lepton flavor violating constraints.
By using the above formulas for the lepton flavor violating processes 
and the neutrino oscillation condition (\ref{c-oscil}), the branching 
ratio of $\mu\rightarrow e\gamma$ and $\tau\rightarrow \mu\gamma$ 
predicted by the model can be plotted in the $(M_1,~\mu_\eta)$ plane if
$M_3$, $m_0$ and $\bar\lambda$ are settled.
It is useful to note that these bounds become more severe by making
$\bar\lambda$ smaller generally since a smaller $\bar\lambda$ requires
larger values for $|h_i|$ as found from eqs.~(\ref{nmass}) and (\ref{mscale}).
We fix these values to
\begin{equation}
 M_3=7.8~{\rm TeV}, \quad m_0=0.48~{\rm TeV}, \quad 
\bar\lambda=1.24\times 10^{-9} 
\end{equation}
as a typical example. The last one corresponds 
to $\lambda_u\lambda_d=10^{-7.7}$ and $\tan\beta=16$, for example. 
For a while, we consider the case $\varphi_1=\varphi_2$ only.

In the left frame of Fig.~3, we show the contours of these branching 
ratios. Red solid lines represent the contours of 
$Br(\mu\rightarrow e\gamma)\times 10^{11}=1.2$ and 0.6 downward, and
blue dotted ones are contours of 
$Br(\tau\rightarrow\mu\gamma)\times 10^8=0.1$ and 0.04 rightward.
The former one is independent of $M_1$ as is clear from the expression 
in eq.~(\ref{eqlfv}). 
$F_2(M^2_3/m_{\eta a}^2)$ becomes smaller for a larger
$M_3$ although the larger $M_3$ makes the neutrino Yukawa coupling 
$h_3$ larger through eq.(\ref{c-oscil}). Thus,
this branching ratio can easily satisfy the present experimental 
bounds by making $M_3$ large enough. 
We note that this feature is intimately related to 
the flavor structure (\ref{yukawa}) which induces the tri-bimaximal 
MNS matrix.
$Br(\tau\rightarrow\mu\gamma)$ is found to show a different 
behavior, which is mainly controlled by the masses 
and the couplings of $\bar N_{1,2}$.
This figure shows that the present experimental bounds for these
processes \cite{clfv} are satisfied at the wide range 
of the $m_{\eta -}>M_1$ regions, where $m_{\eta-}=M_1$ is plotted 
by a thin black dashed line.

In the right frame of Fig.~3, as in the left frame 
each branching ratio is plotted by a red
solid line and a blue dotted line as
a function of $m_0$ for $M_1=2.5$~TeV and $\mu_\eta=4.2$~TeV.
The present bounds for them are also plotted by the thin black solid and
dashed lines, respectively. 
The figure suggests that the constraint from the $\mu\rightarrow e\gamma$
requires $m_0~{^>_\sim}~0.45$~TeV, although the
$\tau\rightarrow\mu\gamma$ gives no constraint on $m_0$.
From these figures, we find that the model can be easily consistent with 
both the neutrino oscillation data and the lepton flavor violating
constraints for the natural values of parameters 
as long as they are suitably fixed.

As a related subject, it is useful to note that the similar diagram 
to Fig.~2 contributes to the electric dipole moment of an electron 
(EDME) and the muon $g-2$.
Even if the neutrino Yukawa couplings $h^N_{ei}$ are complex, the EDME is
not induced since there is no mixing among $\bar N_i$'s. 
New contributions to the muon $g-2$ due to the similar diagram to 
Fig.~2 are summarized as
\begin{eqnarray}
&&\delta a_\mu\simeq \frac{m_\mu^2}{2(4\pi)^2}\sum_{a=\pm}\left[
\frac{-1}{m_{\eta a}^2}\left\{(|h_1|^2+|h_2|^2)F_2
\left(\frac{M_1^2}{m_{\eta a}^2}\right)+
|h_3|^2F_2\left(\frac{M_3^2}{m_{\eta
	   a}^2}\right)\right\}\right. \nonumber \\
&&\hspace*{3.2cm}\left.+\frac{a}{\mu_\eta^2}\left\{(|h_1|^2+|h_2|^2)F_2
\left(\frac{M_{1a}^2}{\mu_\eta^2}\right)+
|h_3|^2F_2\left(\frac{M_{3a}^2}{\mu_\eta^2}\right)\right\}
\right].
\end{eqnarray}
We estimate it for the allowed parameter sets obtained 
in the above analysis. 
The results seem to be smaller by three order of
magnitude in comparison with 
$\delta a_\mu=(30.2\pm 8.7)\times 10^{-10}$, which is a
discrepancy shown by the SM prediction and the value derived by the 
experiment \cite{g2}.
This suggests that another origin is required for the explanation 
of this muon $g-2$ discrepancy.
  
\subsection{Two dark matter candidate}
The model has two types of the DM candidate in general.
One of them is the lightest neutralino $\chi$ whose stability is 
guaranteed by the $R$-parity as in case of the MSSM.  
The other one is the lightest neutral field with the odd parity of the new
$Z_2$ symmetry, which is the remnant symmetry of the anomalous U(1)$_X$.
It corresponds to the lightest neutral state composed of the components of 
the chiral supermultiplets $\bar N_i$, or $\eta_{u,d}^0$ and $\phi$.
In the following study, we assume that the singlet fermion 
$\psi_{N_1}$ (the fermionic component of $\bar N_1$) is the lightest 
one among these candidates. 
Since this $Z_2$ is not an exact symmetry but is weakly broken as shown in
eq.~(\ref{anom}) by the anomaly effect, the latter candidate
$\psi_{N_1}$ is not stable 
but it could have a long lifetime
comparable to the age of the universe.
The condition for this possibility is discussed in the next section.
If this is the case, the DM relic abundance suggested by the WMAP \cite{wmap}
should be satisfied by these two contributions such as
\begin{equation}
\Omega_{\chi}h^2+\Omega_{\psi_{N_1}}h^2=0.11.
\label{wmap}
\end{equation}

\begin{figure}[t]
\begin{center}
\epsfxsize=12cm
\leavevmode
\epsfbox{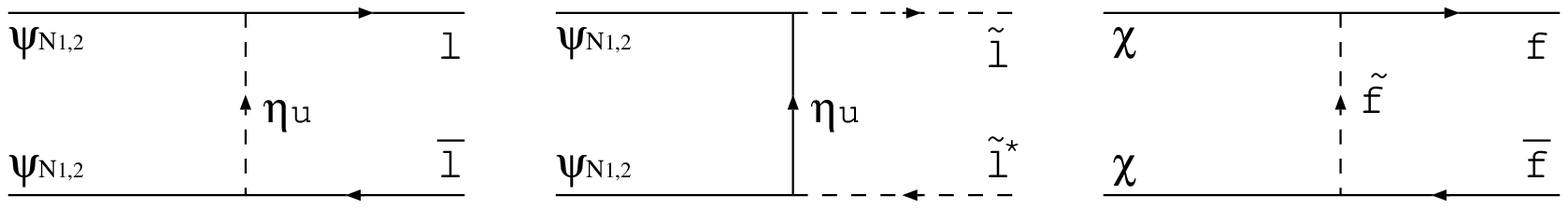}
\end{center}

\vspace*{-3mm}
{\footnotesize {\bf Fig.~4}~~Diagrams contributing to the $t$-channel 
annihilation of the two DM candidates. 
There are also $u$-channel diagrams corresponding to these.
Although the final state $f\bar f$ 
of the $\chi$ annihilation is composed of both $q\bar q$ and 
$\ell\bar\ell$, the $\psi_{N_1}$ annihilation contains
 $\ell\bar\ell$ only.}
\end{figure}

The annihilation of two $\psi_{N_1}$'s is induced through the 
$t$- and $u$-channel $\eta_u$ exchange. 
Final states of such processes are composed of 
a pair of lepton and antilepton or a pair of slepton and antislepton. 
The latter final state gives new contributions compared with 
the non-supersymmetric case 
as long as $M_1>\tilde m_L(\equiv m_0)$ is satisfied for the slepton
mass $\tilde m_L$. 
If $\psi_{N_2}$ has a mass almost degenerate with the one of 
$\psi_{N_1}$ and this is the case here, we need to consider 
the coannihilation effect \cite{coann}. Final states of this annihilation 
are controlled by the flavor structure of the neutrino Yukawa couplings
which is shown in eq.~(\ref{yukawa}). 
On the other hand, the annihilation of two $\chi$'s occurs through various
processes depending on its composition, which is determined by both the
supersymmetry breaking scenario and the radiative symmetry breaking
conditions (see Fig.~4).  
It has been studied in detail in the MSSM context \cite{susydm}.
Here we do not fix the scenario but just confine our study to
the pure bino case by assuming that $\mu$ and the masses of Higgs
doublet and gauginos are tuned to satisfy the required condition, 
for simplicity. 
In this case, the annihilation is expected to occur dominantly 
through $t$- and $u$-channel sfermion 
exchange.\footnote{If the lightest neutralino $\chi$ is heavier than 
Higgs scalars, we need to take account of the $t$-channel Higgsino 
exchange process. However, it is expected to be subdominant as long as
the Higgsino is heavier than the sfermions,
which is assumed throughout this analysis.} 
We estimate the annihilation 
cross section $\sigma v$ for these DM candidates 
by expanding it as $\sigma v=a+bv^2$
in powers of their relative velocity $v$ \cite{griest}. 

For the singlet fermion $\psi_{N_1}$, we need to take account of the
coannihilation effect with $\psi_{N_2}$ because of the assumption
$M_1\simeq M_2$. 
In order to estimate the freeze-out temperature $T_f$ of $\psi_{N_1}$ 
including the coannihilation case, we follow the procedure given in 
\cite{coann}. 
We define $\sigma_{\rm eff}$ and $g_{\rm eff}$ as
\begin{eqnarray}
\sigma_{\rm eff}&=& 
\frac{g_{N_1}^2}{g_{\rm eff}^2}\sigma_{\psi_{N_1}\psi_{N_1}}+
2\frac{g_{N_1}g_{N_2}}{g_{\rm eff}^2}
\sigma_{\psi_{N_1}\psi_{N_2}}(1+\delta)^{3/2}e^{-x\delta}
+\frac{g_{N_2}^2}{g_{\rm eff}^2}\sigma_{\psi_{N_2}\psi_{N_2}}
(1+\delta)^3e^{-2x\delta}, \nonumber \\
g_{\rm eff}&=&g_{N_1}+g_{N_2}(1+\delta)^{3/2}e^{-x\delta},
\label{effcross}
\end{eqnarray}
where internal degrees of freedom of $\bar N_i$ are described by $g_{N_i}$ and  
$\delta\equiv (M_2-M_1)/M_1$. 
If we define $a_{\rm eff}$ and $b_{\rm eff}$ by 
$\sigma_{\rm eff}v=a_{\rm eff}+b_{\rm eff}v^2$, 
the thermally averaged cross section can be written as 
$\langle\sigma_{\rm eff}v\rangle=a_{\rm eff}+ 6 b_{\rm eff}/x$ where 
$x=M_1/T$.
Since $\delta\ll 1$ is supposed here, the second and third terms can bring
the important contribution.
Using these formulas, the effective annihilation cross section is given by
\begin{eqnarray}
(\sigma_{\psi_{N_1}})_{\rm eff}v&\simeq&
\frac{Y_s^4}{8\pi}\sum_{a,b=\pm}
\frac{M_1^2}{(M_1^2+m_{\eta a}^2)(M_1^2+m_{\eta b}^2)}(1+ p_Fv^2)\nonumber \\
&+&\frac{Y_p^4}{96\pi}\sum_{a,b=\pm}
\frac{M_1^2(M_1^4+m_{\eta a}^2m_{\eta b}^2)}
{(M_1^2+m_{\eta a}^2)^2(M_1^2+m_{\eta b}^2)^2}v^2 \nonumber \\
&+&\frac{Y_s^4}{8\pi}
\frac{M_1^2\beta}{(\mu_{\eta}^2+M_1^2\beta^2)^2}\left(\beta^2+p_Sv^2\right)
\nonumber \\
&+&\frac{Y_p^4}{32\pi}
\frac{M_1^2\beta}{(\mu_{\eta}^2+M_1^2\beta^2)^2}\left(1+\frac{2\beta^2}{3}
-\frac{4\mu_\eta^2M_1^2\beta^2}{3(\mu_\eta^2+M_1^2\beta^2)^2}\right) v^2,
\label{effcr}
\end{eqnarray}
where $\beta=\sqrt{1-(m_0/M_1)^2}$.
The mass eigenvalue $m_{\eta a}$ is given in eq.~(\ref{etamass}).  
$Y_s$ and $Y_p$ are defined by
\begin{equation}
Y_s^4=2|h_1|^2|h_2|^2\sin^2(\varphi_1-\varphi_2), \qquad
Y_p^4=|h_1|^4+2|h_1|^2|h_2|^2\cos 2(\varphi_1-\varphi_2)+|h_2|^4.
\end{equation}
The first and second lines in eq.~(\ref{effcr}) represent the contributions 
with the lepton-antilepton final state. The third and forth lines 
come from the slepton-antislepton final state. 
We find that there can be $s$-wave contributions in the
first and third lines if the neutrino Yukawa 
couplings $h^N_{ij}$ have the phases such as 
$\varphi_1\not=\varphi_2+n\pi$.\footnote{The $s$-wave
contributions are dominant in such cases. The expressions for 
$p_{F,S}$ for the corresponding $p$-wave contributions are shown 
in Appendix C for the completeness.} 
This happens since the coannihilation cross section 
$\sigma_{\psi_{N_1}\psi_{N_2}}$ 
allows the $s$-wave contribution \cite{recon}.

\begin{figure}[t]
\begin{center}
\epsfxsize=8cm
\leavevmode
\epsfbox{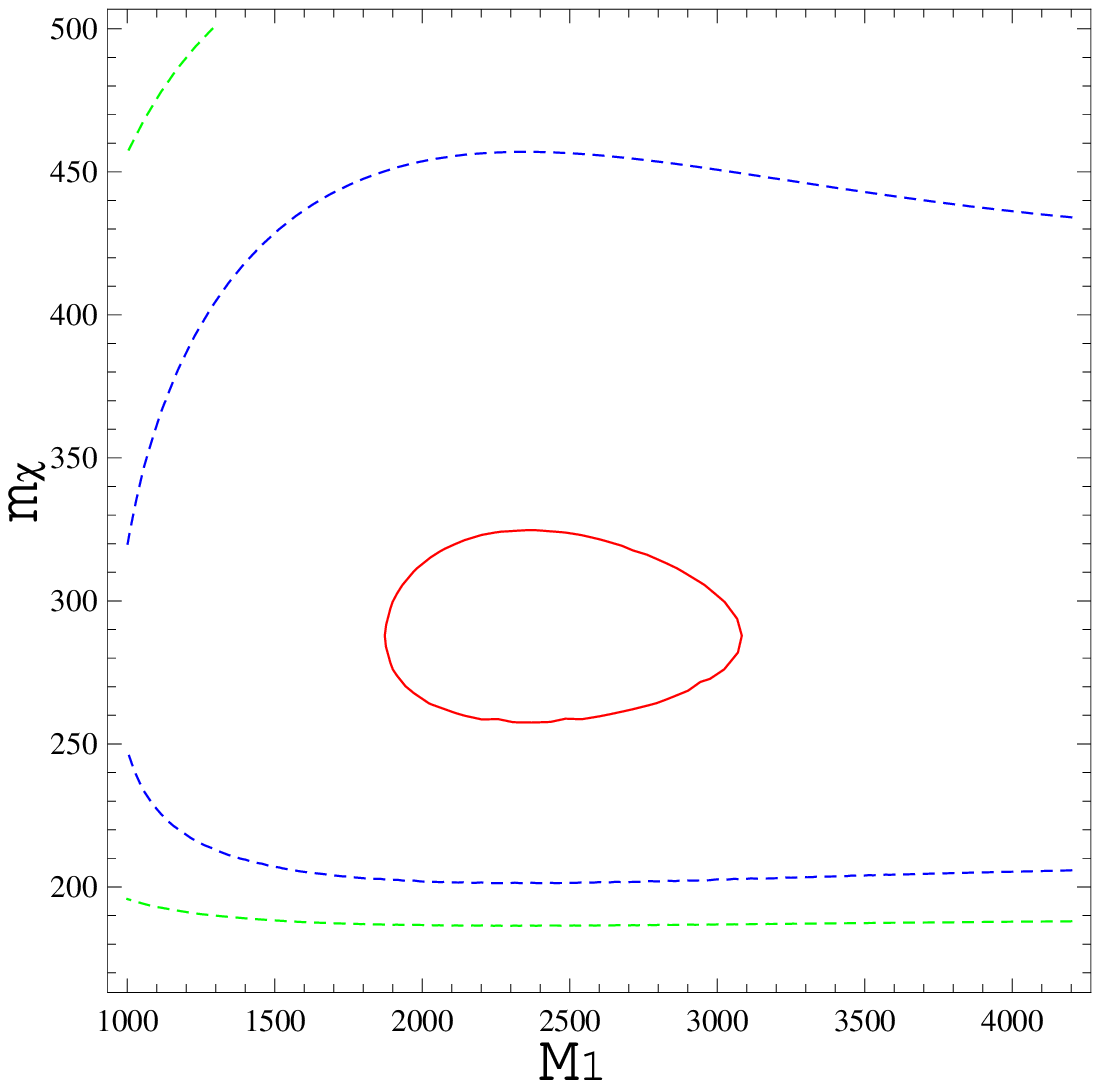}
\hspace*{5mm}
\epsfxsize=6.5cm
\leavevmode
\epsfbox{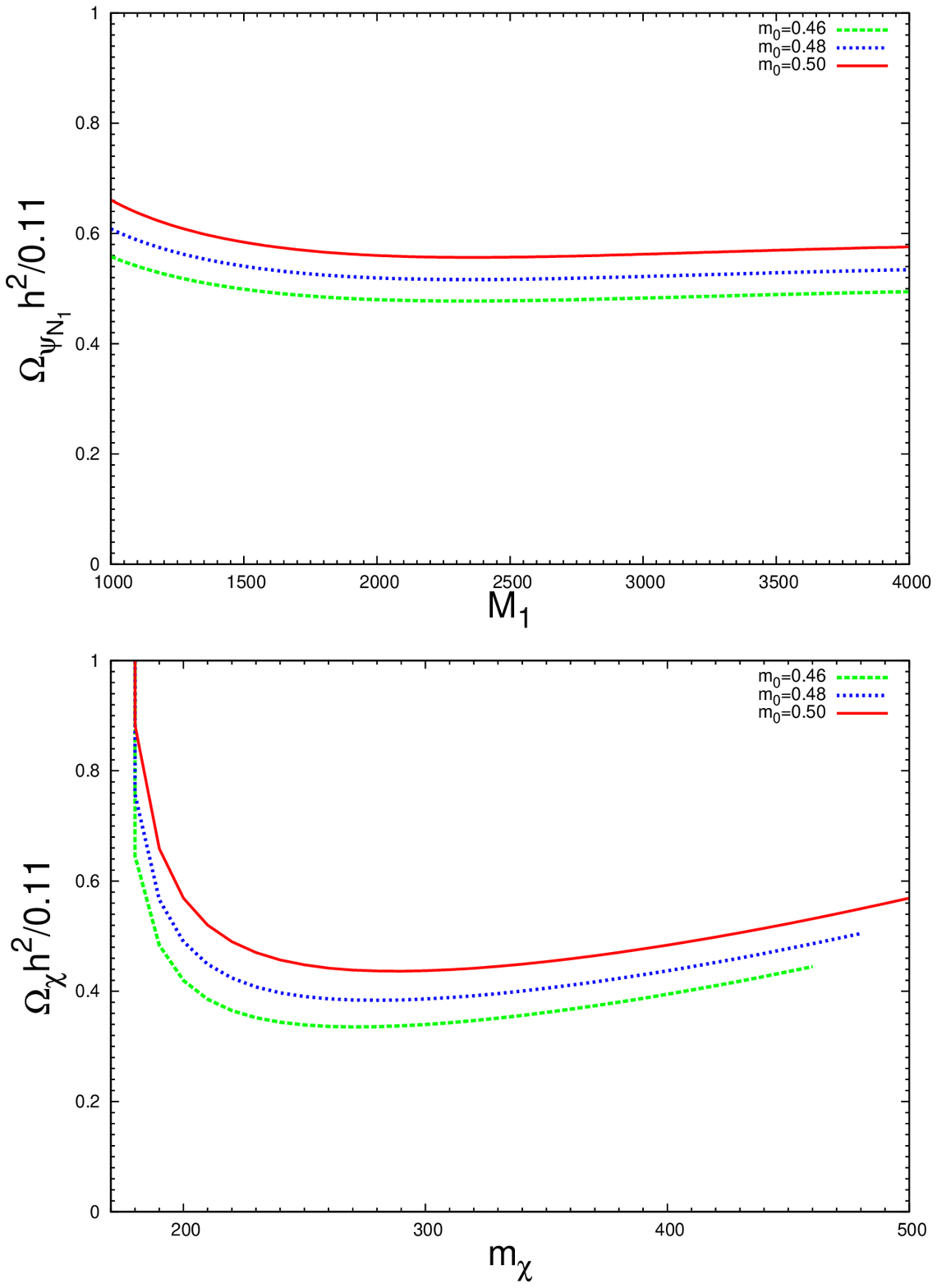}
\end{center}
\vspace*{-3mm}
{\footnotesize {\bf Fig.~5}~~The left frame shows the 
contours of $\Omega_{\psi_{N_1}} h^2+\Omega_\chi h^2=0.11$ for
$m_0=0.46~{\rm TeV}$ (a green dotted line), 0.48~TeV (a blue dotted line), 
and 0.50~TeV (a red solid line). 
Each figure in the right frames shows the ratio of each DM component to
 the total relic abundance $\Omega_{\psi_{N_1}}h^2/0.11$ and
 $\Omega_\chi h^2/0.11$, respectively. Each line corresponds 
to the same value of $m_0$ used in the left frame. 
In these figures $\varphi_1-\varphi_2=0$
 is assumed.}
\end{figure}

On the other hand, the $s$-wave contribution to the annihilation 
cross section of the bino-like $\chi$ to $\bar ff$ is expressed as \cite{griest}
\begin{equation}
\sigma_{\chi\chi} v\simeq\sum_{f}\frac{4c_f}{\pi}
\frac{G_F^2m_\chi^2m_W^4\beta^\prime}
{( m_0^2+\beta^{\prime 2}m_\chi^2)^2}
\tan^4\theta_W\Big((T_{3L}-Q_f)^4+Q_f^4+2(T_{3L}-Q_f)^2Q_f^2\Big)z^2,
\label{chicross}
\end{equation}
where $z=m_f/m_\chi$, $\beta^\prime=(1-z^2)^{1/2}$ and 
$Q_f$ is the electric charge of $f$.
In the summation in eq.~(\ref{chicross}), $f$ runs among quarks
($c_f=3$) and leptons ($c_f=1$) which satisfy $m_f<m_\chi$.
Although $\chi\chi$ can annihilate to a pair of Higgs scalars 
through the Higgsino exchange, we neglect it as a subdominant process.
If the bino mass $M_{\tilde B}$ is almost degenerate with Higgsino mass
$\mu_H$, the coannihilation between $\chi$ and a Higgsino-like neutralino
$\chi^\prime$ brings important effects on the relic abundance of $\chi$.
The relevant cross sections to this coannihilation are given by \cite{griest}
\begin{eqnarray}
\sigma_{\chi\chi^\prime}v&\simeq&\sum_f\frac{2c_f}{\pi}G_F^2m_\chi^2
\beta^\prime \left[\frac{(g^\prime\cos 2\beta/g\sin\beta)m_Wm_f\xi_f}
{[(m_{H_3^0}^2-4m_\chi^2)^2
+m_{H_3^0}^2\Gamma_{H_3^0}^2]^{1/2}}
-\frac{(T_{3L}\tan\theta_W)m_Wm_f\zeta_f}{m_0^2+\beta^{\prime 2}m_\chi^2}
\right]^2, \nonumber \\
\sigma_{\chi^\prime\chi^\prime}v&\simeq&\sum_f\frac{4c_f}{\pi}G_F^2m_\chi^2
\beta^\prime \left[\frac{(\xi_f\cot\beta/2)(m_{\chi^\prime}+\mu_H)m_f}
{[(m_{H_3^0}^2-4m_\chi^2)^2+m_{H_3^0}^2\Gamma_{H_3^0}^2]^{1/2}}
-\frac{m_f^2\zeta_f^2z}{4(m_0^2+\beta^{\prime 2}m_\chi^2)}\right]^2, 
 \end{eqnarray}
where $\xi_f=\cot\beta$ and $\zeta_f=\frac{1}{\sin\beta}$ for $f$ in the up
sector and $\xi_f=\tan\beta$ for and also $\zeta_f=\frac{1}{\cos\beta}$ 
for $f$ in the down sector. 
The effective cross section including the coannihilation can be
determined by the similar formulas to eq.~(\ref{effcross}).
The dominant contributions for it are expected to come from a channel 
with the $t\bar t$ final state if $m_\chi>m_t$ is satisfied.
If we focus our numerical study to the case with 
$m_\chi\simeq m_\chi^\prime$ and 
fix both $\tan\beta$ and $m_{H_3^0}^2$ to suitable values\footnote{In
this study we assume a sufficiently large value for $m_{H_3^0}^2$ such as
$m_{H_3^0}^2\gg 4m_\chi^2$, for simplicity.},
these cross sections are determined by two free parameters $m_0$ and
$m_\chi$. They should satisfy some required conditions.
Since $\chi$ should be lighter than the left-handed sneutrinos,
$m_\chi< m_0$ has to be satisfied. The new contribution to the
$\mu\rightarrow e\gamma$ imposes the lower bound on $m_0$ as found from
Fig.~3. In relation to this constraint it is useful to remind that
a larger $M_3$ allows smaller values for $m_0$.
 
Now we examine the possibility to realize the required relic abundance
by these two DM. 
If we follow the ordinary method given in
\cite{coann,griest}, we can estimate the relic abundance by using 
the results for the effective annihilation cross section given above.
Each relic abundance $\Omega_{\psi_{N_1}}$ and $\Omega_{\chi}$ 
is given by the formulas
\begin{equation}
\Omega h^2=\frac{1.07\times 10^9 x_f}{g_\ast^{1/2}m_{\rm pl}({\rm
 GeV})
(a_{\rm eff}+ 3b_{\rm eff}/x_f)}, \quad
x_f=\ln\frac{0.038 g_{\rm eff} m_{\rm pl}m_{DM}(a_{\rm eff}
+6b_{\rm eff}/x_f)}{g_\ast^{1/2}x_f^{1/2}},
\end{equation}
where $m_{\rm pl}=1.22\times 10^{19}$~GeV and $m_{DM}$ is 
the mass of DM.
$g_{\rm eff}$ stands for the internal degrees of freedom of DM or
the effective degrees of freedom in the coannihilation case.
We can use $g_\ast\simeq 100$ as the relativistic degrees of freedom at
the DM freeze-out temperature $T_f(\equiv m_{\rm DM}/x_f)$ for both
$\psi_{N_1}$ and $\chi$.

The results are shown in Fig.~5 for the parameter set used in the
previous section.  
In the left frame, we plot the contours for the relic 
abundance (\ref{wmap}) in the $(M_1, m_\chi)$ plane 
for $m_0=0.46$~TeV (a green dotted line),
0.48~TeV (a blue dotted line), and 0.50~TeV (a red solid line). 
Since we are considering that $\chi$ is the DM lighter than
$\psi_{N_1}$, $M_1, m_0>m_\chi$ should be satisfied. 
The allowed regions in the $(M_1,m_\chi)$ plane are represented by 
the points on each contour which satisfy this condition.
They are found to have the almost fixed values of $m_\chi$ for the larger
values of $M_1$. 
The reason can be found in the right frame of Fig.~5, where the relic abundance 
of each DM component is plotted for the same parameters as the left
frame. The same lines are used as the ones for 
the corresponding contours in the left frame. 
Since $\Omega_{\psi_{N_1}}$ is almost constant at large $M_1$ regions, 
the condition (\ref{wmap}) can be satisfied only for restricted values
of $\Omega_{\chi}$.  
Since both DM components have the same order abundance, 
we can expect rather different DM phenomenology from the one component 
DM models.

\begin{figure}[t]
\begin{center}
\epsfxsize=7cm
\leavevmode
\epsfbox{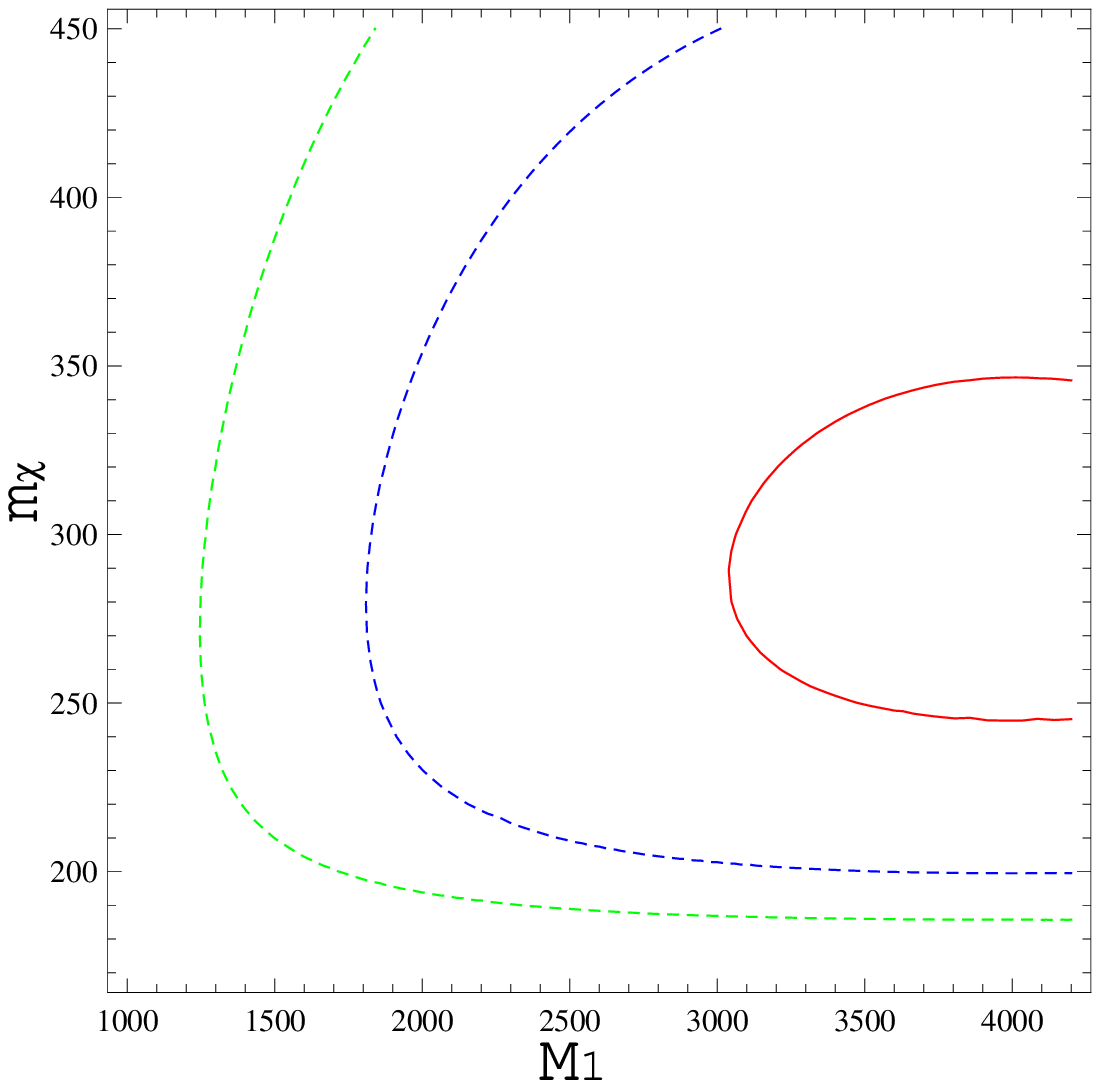}
\hspace*{5mm}
\epsfxsize=7cm
\leavevmode
\epsfbox{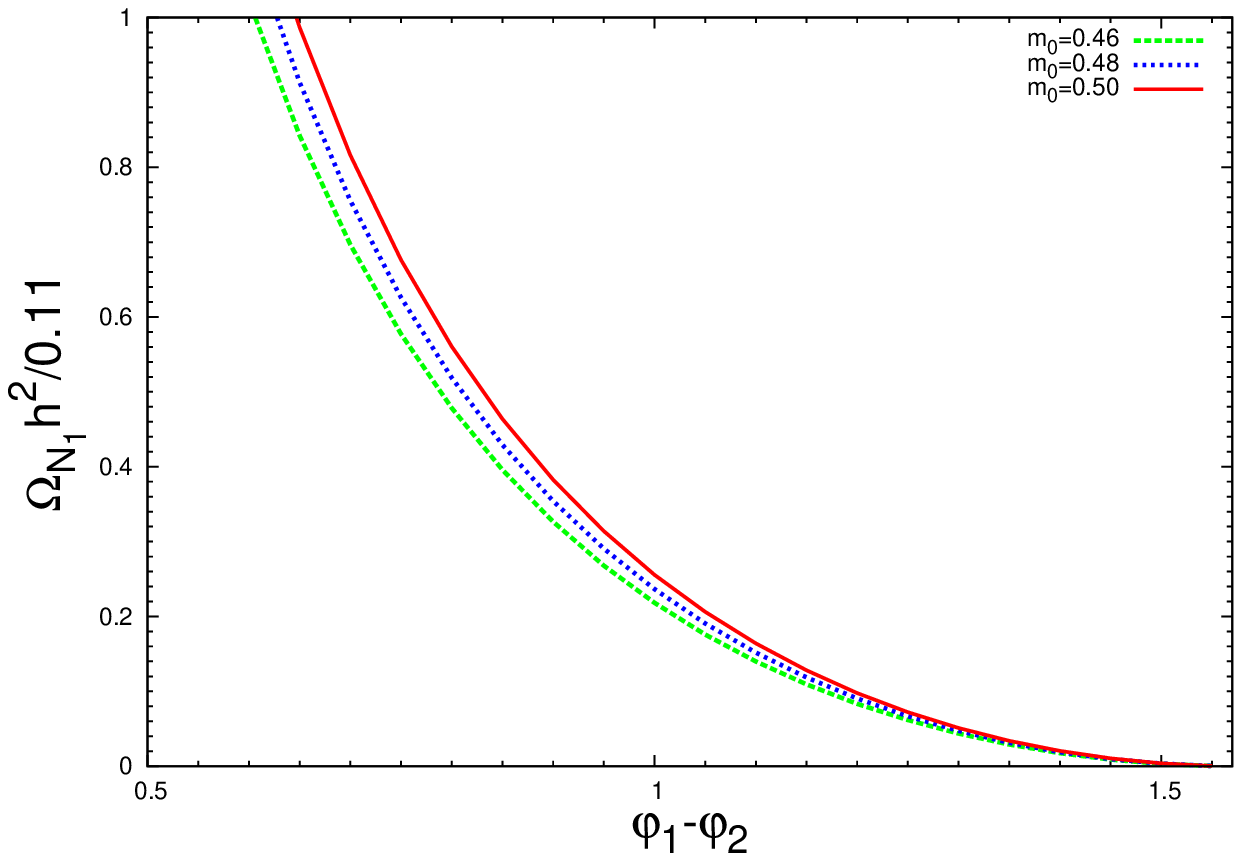}
\end{center}
\vspace*{-3mm}
{\footnotesize {\bf Fig.~6}~~The left frame is the same one as the left
 frame in Fig.~5. However, $\varphi_1-\varphi_2=\pi/4$ is assumed here.
The right frame shows the phase dependence of $\Omega_{N_1}h^2$ for each
 value of $m_0$ in the left frame.}
\end{figure}
   
We need to note that the above solutions are obtained for rather large
neutrino Yukawa couplings such as 
$\sqrt{|h_1|^2+|h_2|^2}\simeq 3.56$ in case of $m_0=0.48$~TeV, for example.
Although these results on the neutrino masses and the DM 
are interesting enough for the model, such large Yukawa couplings 
are dangerous for the perturbatibity and the stability of 
the model \cite{bw-sty}.
However, this point can be improved by considering the case
$\varphi_1-\varphi_2\not=n\pi$, which makes the $s$-wave contributions to the
$\psi_{N_1}$ annihilation cross section possible. Since the $s$-wave
contributions give much larger effects than the $p$-wave ones for the
thermally averaged annihilation cross section $\langle\sigma v\rangle$,
the neutrino Yukawa couplings required to reduce the relic abundance of
$\psi_{N_1}$ can be smaller. In the left frame of Fig.~6, 
the contours corresponding to the ones in the left frame in Fig.~5 
are plotted by assuming $|h_1|=|h_2|$ and $\varphi_1-\varphi_2=\pi/4$.
The parameters are fixed to the same values as in Fig.~5 except for
$\bar\lambda$, which is taken here as $\bar\lambda=8.4\times 10^{-9}$.
Since the neutrino Yukawa couplings have much smaller values 
like $\sqrt{|h_1|^2+|h_2|^2}\simeq 1.45$ in case of $m_0=0.48$~TeV, 
the above mentioned tension is relaxed. 
In the right frame of Fig.~6, the phase dependence of 
$\Omega_{N_1}h^2$ is shown for the same parameter setting. 
 
\section{Probing two dark matter}
The present model has two DM components, that is, the meta stable
lightest singlet fermion $\psi_{N_1}$ and the lightest neutralino $\chi$. 
They are expected to be observed by several kinds of experiments.
The former one may be studied indirectly through the 
decay products such as charged particles and gamma rays. 
The anomaly reported in the cosmic rays by PAMELA and Fermi-LAT may
be relevant to this decay. 
The latter one may be observed directly through the elastic 
scattering with nuclei in the same way as the ordinary lightest
superparticle. However, the situation can be rather different from 
the MSSM since the DM relic abundance is composed 
of two components. In this section these subjects are briefly discussed and
the detailed study will be presented elsewhere.

\subsection{Decay of the right-handed neutrino dark matter}
The relic DM in this model can be composed of the two components
$\psi_{N_1}$ and $\chi$ as shown in the previous part.
However, one of the DM candidates $\psi_{N_1}$ is not stable since
the $Z_2$ symmetry which guarantees its stability is not exact.
It should be reminded that this symmetry is the remnant symmetry left
after the spontaneous breaking of the anomalous U(1)$_X$. 
Green-Schwarz anomaly cancellation mechanism induces
the $Z_2$ violating interaction and also the corresponding soft 
supersymmetry breaking term, which are shown in eqs.~(\ref{anom}) 
and (\ref{softsb}).
If $\psi_{N_1}$ is heavier than $\chi$, this interaction brings 
the decay of $\psi_{N_1}$ to $\chi$ through the diagrams shown in Fig.~7.

\begin{figure}[t]
\begin{center}
\epsfxsize=10cm
\leavevmode
\epsfbox{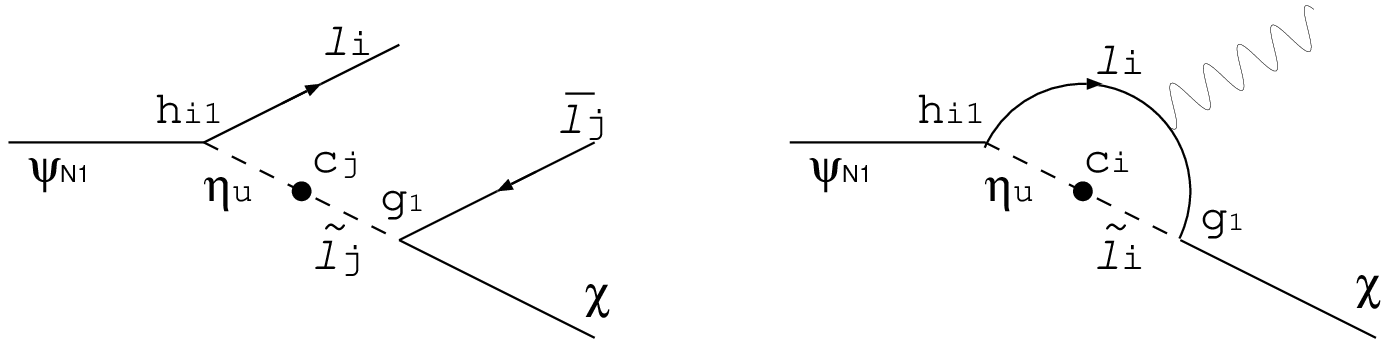}
\end{center}

\vspace*{-3mm}
{\footnotesize {\bf Fig.~7}~~Decay processes of $\psi_{N_1}$ to $\chi$.
A bulb represents the anomaly induced interaction 
$c_jBM_{\rm pl}e^{-b_j}\tilde L_j\tilde\eta_u$. }
\end{figure}

We can estimate the lifetime of $\psi_{N_1}$ due to the decay 
derived by this interaction.
It can be expressed as 
 \begin{eqnarray}
 \tau_{\psi_{N_1}} \sim \left(\frac{ \mbox{1~TeV}}{M_1}\right)^5
 \left( \frac{\mu_\eta}{\mbox{1.3~TeV}} \right)^4
 \left( \frac{ m_0}{\mbox{1~TeV}}\right)^4
 \left(\frac{\mbox{1~TeV}}{B}\right)^2
\left( \frac{e^{2b_j}}{10^{80}}\right)\times 10^{26}  ~~\mbox{sec},
\label{life}
 \end{eqnarray}
where we use $h^N_{i 1}=O(1)$ and $c_j=O(1)$. If $M_1\gg m_0$ is
satisfied, $m_0$ should be replaced by $M_1$ in eq.~(\ref{life}).
From this formula, we find that $\psi_{N_1}$ can have a 
sufficiently long lifetime compared with the age of universe
as long as $b_j >82$ is satisfied.
Thus, although the true stable DM is the lightest neutralino $\chi$, 
we need to take account of the contribution of $\psi_{N_1}$ to the relic 
DM abundance in the universe as discussed in the previous part.

On the other hand, depending on the scale of the $Z_2$ 
breaking $c_jM_{\rm pl}Be^{-b_j}$ in eq.~(\ref{softsb}), 
particles yielded in the decay of 
$\psi_{N_1}$ could bring additional contributions to the cosmic rays in
the present universe.
They could be detected as the anomaly in the expected flux of the cosmic
rays through the various observation. 
In fact, if $b_j\sim 92$ is satisfied, this anomaly induced superweak 
interaction
causes a large enhancement factor of $O(10^{80})$ in eq.~(\ref{life})
to realize a long lifetime of $O(10^{26})$ sec for 
$\psi_{N_1}$.\footnote{It is interesting to note that 
this value of $b_j$ can be consistent with the value of ${\rm Tr}X$ for 
$g_X=O(1)$, which is required to realize the vacuum with the desired 
values of $\varepsilon_\pm$ as discussed below eq.~(\ref{vev}) and in
Appendix A.}
This lifetime is known to be suitable to explain the charged 
cosmic ray anomaly reported by PAMELA \cite{pamela} and Fermi-LAT \cite{fermi}.
Since $\psi_{N_1}$ couples only with leptons 
and sleptons because of the U(1)$_X$ symmetry, 
this decay can yield a pair of a lepton and an antilepton,
or a photon in addition to the lightest neutralino $\chi$. 
This feature is favored by the lepto-philic nature of the PAMELA observations.
Moreover, the flavor structure of the neutrino Yukawa couplings
(\ref{yukawa}) can restrict 
the flavor of the final charged leptons to $\mu$ and $\tau$.\footnote{In
that case we need to impose $c_e=0$ additionally in the anomaly induced
interaction $c_j M_{\rm pl}e^{-b_j}L_j\eta_u$.}
Since this makes positrons and electrons produced by their decay
soft, the model becomes favorable for the explanation of
a plateau at high energy regions of $e^++e^-$ spectrum found in the 
Fermi-LAT observations \cite{fl}.

The heavier DM component $\psi_{N_1}$ has also a radiative decay mode to the
lightest neutralino $\chi$. Its one-loop diagram is shown in Fig.~7.
This decay associates a characteristic gamma which can be
detected through the observation of the cosmic gamma rays.
It is expected to appear as a line shape spectrum at 
the energy $E_\gamma=(M_1^2-m_{\chi}^2)/2M_1$, which corresponds 
to the endpoint of the gamma-ray spectrum caused by the bremsstrahlung
and the inverse Compton scatterings associated to the charged decay
products of $\psi_{N_1}$. This could be a clear evidence of the model.

\begin{figure}[t]
\begin{center}
\epsfxsize=10cm
\leavevmode
\epsfbox{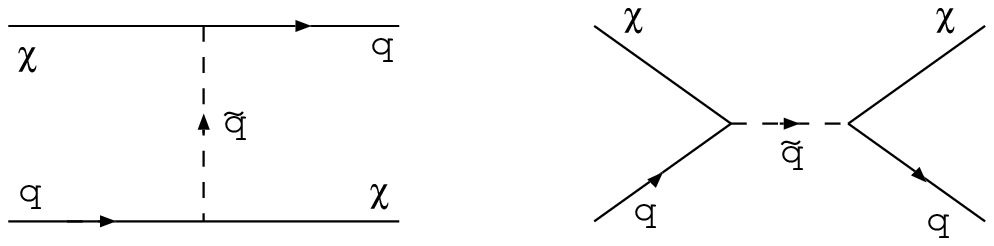}
\vspace*{-3mm}
\end{center}

{\footnotesize {\bf Fig.~8}~~Elastic scattering diagrams
of the bino-like neutralino $\chi$ with the quark
 which are relevant to the direct search of $\chi$.}
\end{figure}

\subsection{Direct detection of the neutralino dark matter}
Direct detection of the DM is expected to 
clarify the nature of DM \cite{susydm}.
Several experiments such as CDMSII, XENON100 and XMASS to observe 
its elastic scattering with nuclei are now under going or will 
start in the near future. 
In the study of DM models, it is crucial to address the discriminative
features of the model, which are expected to be shown in these experiments. 

Since one DM component $\psi_{N_1}$ interacts with the leptons only and 
can not have interactions with nuclei at tree level, 
the scattering cross section with nuclei is heavily suppressed 
by the loop factor. 
Thus, it is difficult to detect it in these experiments.
On the other hand, the neutralino DM $\chi$ can be scattered with nuclei 
at tree level since it has the same nature as an ordinary neutralino 
in the MSSM. One may consider that there is no distinction with
the MSSM case. 
However, it should be noted that the detection rate $R$ in this model can
be different from the direct detection rate $R_{\rm MSSM}$ 
in the MSSM even if the DM $\chi$ has the same mass and 
the same scattering cross section with a nucleon in two models.
The detection rate is approximately defined by
\begin{equation}
R\simeq \sum_i n_i \frac{\rho_\chi}{m_\chi}\langle \sigma_{i\chi}\rangle,
\end{equation}
where $n_i$ is the number of $i$ nuclei species in the detector and 
$\langle \sigma_{i\chi}\rangle$ is the scattering cross section of
the $\chi$ and $i$ nuclei species averaged over the relative velocity
between the $\chi$ and the detector.
Since the present model has two DM
components, the detection rate in this model is related to the MSSM one as   
\begin{equation}
R=\frac{\Omega_\chi}{\Omega_{\psi_{N_1}}+\Omega_\chi}R_{\rm MSSM},
\label{rate}
\end{equation}
for the fixed $m_\chi$ and $m_0$.
Therefore, in the present model the parameter regions to realize 
the same DM detection rate can be changed from the ones in the MSSM, 
although the interactions of the $\chi$ with quarks are same as the 
MSSM neutralino. This might open a new possibility for the supersymmetry
breaking parameters, which is considered not to be allowed in the 
MSSM case.

\begin{figure}[t]
\begin{center}
\epsfxsize=8cm
\leavevmode
\epsfbox{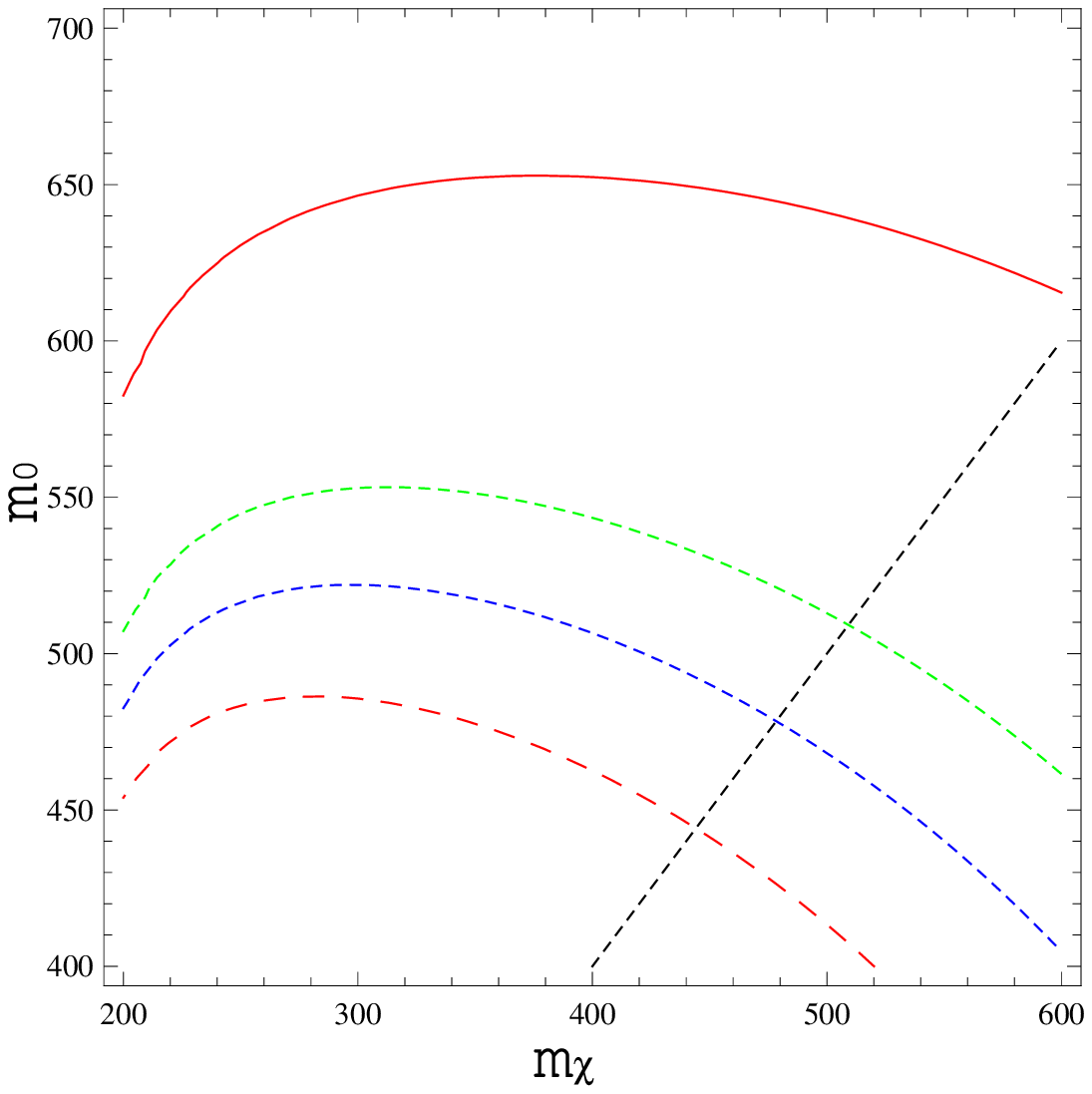}
\end{center}

\vspace*{-3mm}
{\footnotesize {\bf Fig.~9}~~The Contours of $R/R_{\rm MSSM}$ in the
 $(m_\chi,m_0)$ plane. Each line corresponds to
$R/R_{\rm MSSM}=1$ (a red solid line), 0.6 (a green dotted line), 0.5 (a
 blue dotted line) and 0.4 (a red dashed line). 
A thin black dashed line represents a line for $m_0=m_\chi$.}
\end{figure}
 
In order to see this aspect we consider the bino-like $\chi$, as an example.
The dominant contributions to the elastic scattering of the bino-like
$\chi$ with nuclei come from the squark exchange shown in Fig.~8. 
In the MSSM case, the allowed parameter regions relevant to the
detection rate have been extensively studied.
We are interested here in possible changes of the detection rate
from the MSSM and also the change of the allowed region of the
supersymmetry breaking parameters.

We suppose the values used in Fig.~5 for the parameters relevant to the
neutrino mass generation.
Eq.~(\ref{rate}) shows that the figure of $R/R_{\rm MSSM}$ as a 
function of $m_\chi$ is the same as the one in the right frames of
Fig.~5 for the case $\varphi_1-\varphi_2=0$. 
If we considered the case with $m_0=0.48$~TeV and $M_1=2.5$~TeV as an example, 
Fig.~5 shows that $m_\chi$ should be 0.2 TeV or 0.45 TeV to satisfy the WMAP
constraint and then $R/R_{\rm MSSM}\simeq 0.5$ or 0.47 for each
$m_\chi$ value, respectively. 
These suppressed detection rates compared with the MSSM are 
realized for much smaller $m_0$ values than the ones in the MSSM.   
The values of $(m_\chi, m_0)$ found through the direct
search of the lightest neutralino are placed on the contours of 
$\Omega_\chi h^2$, which exist in the region where 
$\Omega_\chi h^2<0.11$, $m_0>m_\chi$ and $m_0>0.45$~TeV. The last condition 
comes from the $\mu\rightarrow e\gamma$ constraint in Fig.~3. 
We note that the model could bring new possibilities 
for the detection rate of the neutralino DM and values of $(m_\chi,
m_0)$ relevant to the supersymmetry breaking even if we impose the DM
relic abundance constraint (see Fig.~9).
    
\section{Summary}
We have studied a supersymmetric model with an anomalous U(1) gauge
symmetry from a view point of the neutrino masses and the DM.
The model considered in this paper may be recognized as a supersymmetric 
extension of the radiative seesaw model for the neutrino masses
proposed by Ma.
The Froggatt-Nielsen mechanism based on the spontaneously broken
anomalous U(1) symmetry generates the hierarchical structure of Yukawa
couplings of the quarks and the charged leptons. Moreover, it can also 
explains the hierarchical couplings and mass scales required for the radiative 
generation of the neutrino masses. If we assume a flavor structure for
the neutrino Yukawa couplings, the tri-bimaximal mixing is automatically
induced. 

The model has two DM components. One is stable and
the other is the decaying DM. Their stability is guaranteed by two
$Z_2$ symmetries, one of which is the ordinary $R$ parity. 
Since one of these discrete symmetries is assumed to be anomalous and then
the $Z_2$ violating interaction is generated nonperturbatively, 
the instability of one DM component is caused.  
This phenomenon happens since this $Z_2$ symmetry is
embedded into the anomalous U(1) gauge symmetry. 
Since the $Z_2$ violating nonpertubative interaction is extremely weak,
the huge suppression factor for the decay width of this DM can be derived.
As a result, its lifetime becomes longer than the age of the universe.
These DM particles can be detected through both direct and indirect
searches. Thus, the model may be checked through these experiments. 
In particular, the recent and future data coming from the cosmic ray 
observations might clarify the relation between this model and the physics
beyond the SM. 
\vspace*{10mm}

\section*{Acknowledgement}
The authors thank Mr.T.~Yoshida for the collaboration at the first 
stage of the work.
This work is partially supported by a Grant-in-Aid for Scientific
Research (C) from Japan Society for Promotion of Science (No.21540262)
and also a Grant-in-Aid for Scientific Research on Priority Areas 
from The Ministry of Education, Culture, Sports, Science and Technology 
(No.22011003).
 
\newpage
\noindent 
{\Large\bf Appendix A}
\vspace*{3mm}

\noindent
In this appendix we give two examples of the U(1)$_X$ charge assignment 
which induce the favorable effective parameters for the generation of
the mass eigenvalues and mixings of the quarks and the 
leptons including the neutrinos. If we define $\varepsilon_\pm$ as 
$\varepsilon_\pm=\langle\Sigma_\pm\rangle/M_{\rm pl}$,
eq.~(\ref{effp}) gives the expression for each parameter in the
superpotential $W$ such as  
\begin{eqnarray}
&&h_{ij}^U=y_{ij}^{U}~\varepsilon_-^{n_{Q_i}+n_{U_j}+n_{H_u}}, \quad
h_{ij}^D=y_{ij}^{D}~\varepsilon_-^{n_{Q_i}+n_{D_j}+n_{H_d}}, \quad
h_{ij}^E=y_{ij}^{E}~\varepsilon_-^{n_{L_i}+n_{E_j}+n_{H_d}}, \nonumber \\
&&\mu_H=\lambda_HM_{\rm pl}~\varepsilon_+^{-\frac{n_{H_u}+n_{H_d}}{n_+}},
\label{mssmp}
\end{eqnarray}
for the ones belonging to the MSSM. Other ones are also expressed as
\begin{eqnarray}
&&h_{ij}^N=y_{ij}^N~\varepsilon_-^{n_{L_i}+n_{N_j}+n_{\eta_u}+1}, \quad
\lambda_u=y_{\eta_u}~\varepsilon_+^{-\frac{n_{\eta_u}+n_\phi+n_{H_d}+1}{n_+}},
\quad 
\lambda_d=y_{\eta_d}~\varepsilon_+^{-\frac{n_{\eta_d}
+n_\phi+n_{H_u}+1}{n_+}}, \nonumber \\
&&\mu_\eta=y_{\eta}M_{\rm pl}~\varepsilon_+^{-\frac{n_{\eta_u}
+n_{\eta_d}+1}{n_+}}, \quad
M_{ij}=y_{N_iN_j}M_{\rm pl}~\varepsilon_+^{-\frac{n_{N_i}+n_{N_j}+1}{n_+}},
\quad
\mu_\phi=y_\phi M_{\rm pl}~\varepsilon_-^{2n_\phi+1}.
\label{fnco}
\end{eqnarray} 
In these formulas, it may be natural to suppose that the original 
coupling constants in the nonrenormalizable interactions, that is, 
$y_{ij}^U$, $\lambda_H$ and so on, have values of $O(1)$.

We assume that $\Sigma_-$ and $\Sigma_+$ obtain the VEVs defined by 
$\varepsilon_-\simeq 10^{-1}$ and $\varepsilon_+\simeq 10^{-4}$, respectively.
The U(1)$_X$ charge is assigned to each field as follows,
\begin{equation}
{\rm example~(i)}\quad\begin{array}{l}
n_Q=(6, 5, 3), \quad n_U=(6,5,3), \quad n_D(4,3,3), \nonumber \\
n_L=(9,9,9), \quad n_E=(1,-1,-3),\quad  n_{H_u}=n_{H_d}=-6, \nonumber \\ 
n_N=(-5,-5,-5), \quad n_{\eta_u}=n_{\eta_d}=-5, \quad n_\phi=7, \quad n_+=3. 
\end{array}
\end{equation}
In this case, it is easily found that the mass matrices for the up- and
down-type quarks and the charged leptons take the following form:
\begin{eqnarray}
&&M_U=\left(\begin{array}{ccc}
\varepsilon_-^6 &\varepsilon_-^5 &\varepsilon_-^3 \\ 
\varepsilon_-^5 &\varepsilon_-^4 &\varepsilon_-^2 \\ 
\varepsilon_-^3 &\varepsilon_-^2 &  1 \\ 
\end{array}\right)\langle H_u^0\rangle, \qquad
M_D=\left(\begin{array}{ccc}
\varepsilon_-^4 &\varepsilon_-^3 &\varepsilon_- \\ 
\varepsilon_-^3 &\varepsilon_-^2 &  1 \\ 
\varepsilon_-^3 &\varepsilon_-^2 &  1 \\ 
\end{array}\right)\langle H_d^0\rangle, \nonumber \\
&&M_E=\left(\begin{array}{ccc}
\varepsilon_-^4 & 0 & 0 \\ 
0 &\varepsilon_-^2 &  0 \\ 
0 &  0 &  1 \\ 
\end{array}\right)\langle H_d^0\rangle,
\end{eqnarray}
where these $M_f$ are defined as $\bar\psi_R M_f\psi_L$.
In the charged lepton mass matrix, the off-diagonal couplings 
are supposed to satisfy $y^E_{ij}=0$ (for $i\not=j$). 
From these mass matrices we obtain both the mass eigenvalues and the CKM
matrix in the quark sector as
\begin{eqnarray}
&& m_u:m_c:m_t=\varepsilon_-^6:\varepsilon_-^4:1, \qquad
m_d:m_s:m_b=\varepsilon_-^4:\varepsilon_-^2:1, \nonumber \\
&& V_{us}\sim \varepsilon_-, \qquad V_{ub}\sim \varepsilon_-^3, \qquad
 V_{cb}\sim \varepsilon_-^2. 
\end{eqnarray}
The charged lepton mass eigenvalues satisfy
\begin{equation}
m_e:m_\mu:m_\tau=\varepsilon_-^4:\varepsilon_-^2:1.
\end{equation} 
These can give qualitatively good results as long as
$\varepsilon_-$ takes a value of Cabibbo mixing angle 0.22.
The effective neutrino Yukawa couplings have no
hierarchical structure $h_{ij}^N=y^N_{ij}=O(1)$ as supposed in the text.
Since the effective $\mu_H$ term is generated as $M_{\rm pl}\varepsilon_+^4$,
it can take an appropriate value for the electroweak
symmetry breaking.

Other effective parameters are estimated as
\begin{equation}
\lambda_u\sim\lambda_d=O(\varepsilon_+), \quad 
M_i=O(y_{N_iN_i}M_{\rm pl}\varepsilon_+^3), \quad 
\mu_\eta=O(y_\eta M_{\rm pl}\varepsilon_+^3), \quad 
\mu_\phi=O(M_{\rm pl}\varepsilon_-^{15}), 
\end{equation} 
where we suppose that the off-diagonal couplings $y_{N_iN_j}$ 
are zero.
These parameters are intimately related to the neutrino mass generation 
in the present model.
The values of these parameters used in the text can be realized 
if $y_{N_iN_i}$ and $y_\eta$ have suppressed values of $O(10^{-2})$.

In the above example, 
we implicitly assume that each parameter is
determined by either $\varepsilon_-$ or $\varepsilon_+$ only 
but is not determined by both of them. 
In the next example, we consider that some
of them are determined by both $\varepsilon_-$ and $\varepsilon_+$.
We assume that $\Sigma_-$ and $\Sigma_+$ obtain the VEVs defined by
$\varepsilon_-\simeq 10^{-1}$ and $\varepsilon_+\simeq 4\times 10^{-4}$ again.
The U(1)$_X$ charge is assigned to each field as follows,
\begin{equation}
{\rm example~(ii)}\quad
\begin{array}{l}
n_Q=(6, 5, 3), \quad n_U=(3,2,0), \quad n_D=(5,4,4), \\
n_L=(7,7,7), \quad n_E=(4,2,0),\quad  n_{H_u}=-3,\quad n_{H_d}=-7, \\ 
n_N=(-6,-6,-6), \quad n_{\eta_u}=-2, \quad n_{\eta_d}=-9, 
\quad n_\phi=6, \quad n_+=3. 
\end{array}
\end{equation}
Using these charges, we can easily find what kind of factor determined by 
$\varepsilon_-$ and $\varepsilon_+$ should appear as the lowest order one
for each term in $W$. 
The mass matrices of quarks and charged leptons and also
the neutrino Yukawa couplings $h_{ij}^N$ show the same features as the ones
in the previous example.

The remaining effective parameters are estimated as
\begin{eqnarray}
&&\mu_H=O(M_{\rm pl}\varepsilon_+^4\varepsilon_-^2), \quad
\lambda_u=O(\varepsilon_+\varepsilon_-), \quad
\lambda_d=O(\varepsilon_+^2\varepsilon_-), \nonumber \\
&&M_i=O(M_{\rm pl}\varepsilon_+^3\varepsilon_-), \quad 
\mu_\eta=O(M_{\rm pl}\varepsilon_+^4\varepsilon_-^2), \quad 
\mu_\phi=O(M_{\rm pl}\varepsilon_-^{13}). 
\end{eqnarray}
These also result in the favorable values for the parameters
relevant to neutrino mass generation and also the electroweak symmetry
breaking. However, it is useful to note a following point. 
Since $\lambda_u\gg\lambda_d$ is satisfied in this case, 
the formula for the dominant contribution to the neutrino masses can be
changed. In fact, the term with $\lambda_u^2$ in eq.~(\ref{nmtr2}) 
could cause the similar order contribution as eq.~(\ref{nmtr}) or larger
contribution even if there is a suppression due to the large $\mu_\phi$.  
\vspace*{5mm}

\noindent 
{\Large\bf Appendix B}
\vspace*{3mm}

\noindent
In this Appendix we give the formulas for the contributions 
to the neutrino masses due to the diagrams (b) and (c) in Fig.~1.
These diagrams have the scalar component and the fermion component 
of the chiral superfield $\phi$ as an internal line, respectively.
The expressions for their contributions can be derived by
taking account of eq.~(\ref{etamass}).

The neutrino mass matrix induced by the diagram (b) 
can be expressed as
\begin{eqnarray}
&&({\cal M}_\nu)_{\alpha\beta}=\sum_{a=\pm}\sum_{i=1}^3
\frac{h_{\alpha i}h_{\beta i}M_i(A+a\mu_\phi)^2a}{(4\pi)^2}  \nonumber \\
&&\times\left[\frac{\lambda_d^2v_u^2}{8}
\Big(I(m_{\eta +},m_{\eta +},M_i,m_{\phi a})
+I(m_{\eta -},m_{\eta -},M_i, m_{\phi a})
-2I(m_{\eta +},m_{\eta -},M_i,m_{\phi_a})\Big)\right. \nonumber \\
&&+\frac{\lambda_u^{2}v_d^2}{8}
\Big(I(m_{\eta +}, m_{\eta +}, M_i, m_{\phi a})
+I(m_{\eta -},m_{\eta -},M_i, m_{\phi a})
+2I(m_{\eta +},m_{\eta -},M_i, m_{\phi a})\Big) \nonumber \\
&&\left.+\frac{\lambda_u\lambda_dv_uv_d a}{4}
\Big(I(m_{\eta +},m_{\eta +},M_i, m_{\phi a})
-I(m_{\eta -},m_{\eta -},M_i, m_{\phi a})\Big)\right], 
\label{nmtr2}
\end{eqnarray}
where the function $I$ is defined as
\begin{eqnarray}
I(m_a,m_b,m_c,m_d)&=&\frac{m_a^2~\ln m_a^2}
{(m_b^2-m_a^2)(m_c^2-m_a^2)(m_d^2-m_a^2)}+
\frac{m_b^2~\ln m_b^2}
{(m_c^2-m_b^2)(m_d^2-m_b^2)(m_a^2-m_b^2)} \nonumber\\
&+&\frac{m_c^2~\ln m_c^2}
{(m_d^2-m_c^2)(m_a^2-m_c^2)(m_b^2-m_c^2)}+
\frac{m_d^2~\ln m_d^2}
{(m_a^2-m_d^2)(m_b^2-m_d^2)(m_c^2-m_d^2)}, \nonumber \\
I(m_a,m_a,m_c,m_d)&=&\frac{(m_a^4-m_c^2m_d^2)~\ln m_a^2}
{(m_c^2-m_a^2)^2(m_d^2-m_a^2)^2}+
\frac{m_c^2~\ln m_c^2}
{(m_d^2-m_c^2)(m_a^2-m_c^2)^2} \nonumber\\
&+&\frac{m_d^2~\ln m_d^2}
{(m_c^2-m_d^2)(m_a^2-m_d^2)^2}-
\frac{1}{(m_c^2-m_a^2)(m_d^2-m_a^2)}.
 \label{mnu2}
\end{eqnarray} 

On the other hand, the neutrino mass matrix induced by 
the diagram (c) is estimated as
\begin{eqnarray}
({\cal M}_\nu)_{\alpha\beta}&=&
\frac{\lambda_d^2v_u^2\mu_\eta^2\mu_\phi}{2(4\pi)^2 }
\sum_{a=\pm}\sum_{i=1}^3h_{\alpha i}h_{\beta i}
I(\mu_\eta,\mu_\eta, M_{ia}, \mu_\phi) \nonumber \\
&-&\frac{\lambda_u^{2}v_d^2\mu_\phi}{(4\pi)^2 }
\sum_{a=\pm}\sum_{i=1}^3h_{\alpha i}h_{\beta i}
J(\mu_\eta,M_{ia},\mu_\phi),
\label{nmtr1}
\end{eqnarray}
where $J$ is defined as
\begin{eqnarray}
J(m_a,m_b,m_c)&=&\left(-\frac{m_a^2}{(m_b^2-m_a^2)(m_c^2-m_a^2)}
+\frac{m_a^4(2m_a^2-m_b^2-m_c^2)}{2(m_b^2-m_a^2)^2(m_c^2-m_a^2)^2}\right)
\ln m_a^2 \nonumber \\
&+&\frac{m_b^4~\ln m_b^2}
{2(m_c^2-m_b^2)(m_a^2-m_b^2)^2}
+\frac{m_c^4~\ln m_c^2}
{2(m_b^2-m_c^2)(m_a^2-m_c^2)^2} \nonumber \\
&-&\frac{m_a^2}{2(m_b^2-m_a^2)(m_c^2-m_a^2)}.
\label{mnu1}
\end{eqnarray}

As long as $\mu_\phi \gg \mu_\eta, M_i, m_0, A$ and 
$\lambda_u\simeq\lambda_d$ are satisfied,
these contributions are found to be subdominant in comparison with the 
one induced by the diagram (a) except for the ones proportional to $\mu_\phi^2$
in eq.~(\ref{nmtr2}). These cause non-negligible contributions to 
eq.~(\ref{nmtr}), which are estimated as
\begin{eqnarray}
&&\sum_{i=1}^3
\frac{h_{\alpha i}h_{\beta i}M_i}{(4\pi)^2} 
\left[\frac{\lambda_d^2v_u^2}{4}
\Big(\tilde I(m_{\eta +}, m_{\eta +},M_i)
+\tilde I(m_{\eta -}, m_{\eta -}, M_i)
-2\tilde I(m_{\eta +}, m_{\eta -}, M_i)\Big)\right. \nonumber \\
&&\hspace*{1cm}\left.+\frac{\lambda_u^{ 2}v_d^2}{4}
\Big(\tilde I(m_{\eta +},m_{\eta +}, M_i)
+\tilde I(m_{\eta -},m_{\eta -}, M_i)
+2\tilde I(m_{\eta +},m_{\eta -}, M_i)\Big)\right], 
\label{nmtr3}
\end{eqnarray}
where the function $\tilde I$ is defined as
\begin{eqnarray}
\tilde I(m_a,m_b,m_c)&=&\frac{m_a^2~\ln m_a^2}{(m_b^2-m_a^2)(m_c^2-m_a^2)}
+\frac{m_b^2~\ln m_b^2}{(m_c^2-m_b^2)(m_a^2-m_b^2)}
+\frac{m_c^2~\ln m_c^2}{(m_b^2-m_c^2)(m_a^2-m_c^2)},\nonumber \\ 
\tilde I(m_a,m_a,m_c)&=&\frac{m_a^2-m_c^2+m_c^2~\ln(m_c^2/m_a^2)}
{(m_a^2-m_c^2)^2}.
 \label{mnu3}
\end{eqnarray} 
This could be the same order contribution to the neutrino mass matrix 
as the one given in eq.(\ref{nmtr}). However, it is useful to note that these
contributions do not change the structure of the MNS matrix but 
they only change the values of $\Lambda_i$ somewhat.

\vspace*{5mm}

\noindent 
{\Large\bf Appendix C}
\vspace*{3mm}

\noindent
The $p$-wave contributions in case of $\varphi_1\not=\varphi_2+n\pi$ are
given by
\begin{eqnarray}
&&p_F=\frac{1}{3}-\frac{M_1^2(3M_1^2+5m_{\eta a}^2)}{6(M_1^2+m_{\eta a}^2)^2}-
\frac{M_1^2(3M_1^2+5m_{\eta b}^2)}{6(M_1^2+m_{\eta b}^2)^2}+
\frac{M_1^4}{3(M_1^2+m_{\eta a}^2)(M_1^2+m_{\eta b}^2)}, \nonumber \\
&&p_S=-\frac{1}{2}+\frac{5}{12}\beta^2+
\frac{M_1^2\beta^2(\mu_\eta^2+4M_1^2\beta^2)}
{3(\mu_\eta^2+M_1^2\beta^2)^2}.
\end{eqnarray}

\newpage
\bibliographystyle{unsrt}

\end{document}